\documentclass[useAMS,usenatbib]{mn2e}
\usepackage[english]{babel}
\usepackage{graphicx}
\usepackage{hyperref}
\usepackage{color}
\usepackage{amssymb,amsmath}
\usepackage{hhline}
\title[Photo-z Quality Cuts and their Effect on the Measured Galaxy Clustering]{Photo-z Quality Cuts and their Effect on the Measured Galaxy Clustering}
\author[P. Mart\'{\i}, R. Miquel, A. Bauer, E. Gazta\~{n}aga]{Pol Mart\'{\i}$^{1}$, Ramon Miquel$^{1,2}$, Anne Bauer$^{3}$, Enrique Gazta\~{n}aga$^{3}$\\
$^{1}$Institut de F\'{i}sica d'Altes Energies, Universitat Aut\`{o}noma de Barcelona, E-08193 Bellaterra (Barcelona), Spain\\
$^{2}$Instituci\'o Catalana de Recerca i Estudis Avan\c{c}ats, E-08010 Barcelona, Spain\\
$^{3}$Institut de Ci\`encies de l'Espai (ICE, IEEC/CSIC), E-08193 Bellaterra(Barcelona), Spain}

\begin{document}

\renewcommand{\refname}{REFERENCES}

\pagerange{\pageref{firstpage}--\pageref{lastpage}} \pubyear{2013}

\maketitle

\label{firstpage}

\begin{abstract}
Photometric galaxy surveys are an essential tool to further our understanding of the large-scale structure of the universe, its matter and energy content and its evolution. These surveys necessitate the determination of the galaxy redshifts using photometric techniques (photo-z). Oftentimes, it is advantageous to remove from the galaxy sample those for which one suspects that the photo-z estimation might be unreliable. In this paper, we show that applying these photo-z quality cuts blindly can grossly bias the measured galaxy correlations within and across photometric redshift bins. We then extend the work of \citet{Ho:2012vy} and \citet{Ross:2011cz} to develop a simple and effective method to correct for this using the data themselves. Finally, we apply the method to the Mega-Z catalog, containing about a million luminous red galaxies in the redshift range $0.45 \lesssim z \lesssim 0.65$. 
After splitting the sample into four $\Delta z = 0.05$ photo-z bins using the BPZ algorithm, we see how our corrections bring the measured galaxy auto- and cross-correlations into agreement with expectations. We then look for the BAO feature in the four bins, with and without applying the photo-z quality cuts, and find a broad agreement between the BAO scales extracted in both cases.
Intriguingly, we observe a correlation between galaxy density and photo-z quality even before any photo-z quality cuts are applied. This may be due to uncorrected observational effects that result in correlated gradients across the sky of the galaxy density and the galaxy photo-z precision. Our correction procedure could also help to mitigate some of these systematic effects.
\end{abstract}

\section{Introduction}

During the last decades, galaxy surveys (SDSS, \citet{York:2000gk}; PanSTARRS, \citet{2000PASP..112..768K}; 2dF, \citet{Colless:2001gk}; LSST, \citet{Tyson:2002nh}; VVDS, \citet{LeFevre:2004hv};  WiggleZ, \citet{Drinkwater:2009sd}; BOSS, \citet{2013AJ....145...10D}) have become crucial tools for our understanding of the geometry, content and destiny of the universe. Spectroscopic surveys provide 3D images of the galaxy distribution in the near universe, but many suffer from limited depth, incompleteness and selection effects. 
Imaging surveys solve these problems but, on the other hand,
do not provide a true 3D picture of the universe, due to their limited resolution in the galaxy position along the line of sight, which is obtained measuring the galaxy redshift through photometric techniques using a set of broadband filters (but see~\citet{Benitez:2008fs} for alternative ideas). 

There are two main sets of techniques for measuring photometric redshifts (photo-zs): template methods (e.g. Hyperz, \citet{Bolzonella:2000js}; BPZ, \citet{Benitez:1998br} \& \citet{Coe:2006hj}; LePhare, \citet{Ilbert:2006dp}; EAZY, \citet{2008ApJ...686.1503B}), in which the measured broadband galaxy spectral energy distribution (SED) is compared to a set of redshifted templates until a best match is found, thereby determining both the galaxy type and its redshift; and
training methods (e.g. ANNz, \citet{Collister:2003cz}; ArborZ, \citet{Gerdes:2009tw}; TPZ, \citet{2013MNRAS.432.1483C}), in which a set of galaxies for which the redshift is already known is used to train a machine-learning algorithm (an artificial neural network, for example), which is then applied over the galaxy set of interest.
Each technique has its own advantages and disadvantages, whose discussion lies beyond the scope of this paper. 

Most if not all photo-z algorithms provide not only a best estimate for the galaxy redshift but also an estimate of the quality of its determination, be it simply an error estimatation or something more sophisticated like the {\it odds} parameter in the {\tt BPZ} code~\citep{Benitez:1998br}. Applying cuts on the value of this quality parameter, one can clean up the sample from galaxies with an unreliable photo-z determination~\citep{Benitez:1998br}, or even select a smaller sample of galaxies with significantly higher photo-z precision~\citep{Marti:2013}.

However, we will show in this paper that these quality cuts can affect very significantly the observed clustering of galaxies in the sample retained after the cuts, thereby biasing the cosmological information that can be obtained. Therefore, this effect needs to be corrected. Fortunately, this can be readily achieved using a technique similar to the one presented in~\citet{Ho:2012vy} to deal with, among others, the effect of stars contaminating a galaxy sample. 

The outline of the paper is as follows.
Section~\ref{sec:data} discusses the galaxy samples that we use in our study: the Mega-Z photometric galaxy sample ~\citep{Collister:2006qg} and its companion, the 2SLAQ spectroscopic sample ~\citep{Cannon:2006qh}, which we use to characterize the redshift distribution of the galaxies in Mega-Z. We will also describe the photo-z algorithm used ({\tt BPZ}) and the resulting redshift distributions in four photometric redshift bins in Mega-Z. In section~\ref{sec:clustering} we present the measurement of the galaxy-galaxy angular correlations within, and cross-correlations across, the four photo-z bins in Mega-Z,
comparisons with the theoretical expectations, 
and the effects of applying several photo-z quality cuts to the data. Section~\ref{sec:correction} introduces the correction we have devised for the effect of the photo-z quality cut and applies it to the data, resulting in corrected angular correlation functions that we then compare with the predictions. In section~\ref{sec:BAO} we extract the Baryon Acoustic Oscilation (BAO) angular scale from the corrected data and compare it with the result obtained without any photo-z quality cuts.
Finally, in section~\ref{sec:discussion} we discuss the relevance of our results, particularly for previous studies that applied photo-z quality cuts while ignoring their effects on clustering, and we offer some conclusions.

\section{Data Samples}
\label{sec:data}

\begin{figure*}
\centering
\includegraphics[type=pdf,ext=.pdf,read=.pdf, width=130mm]{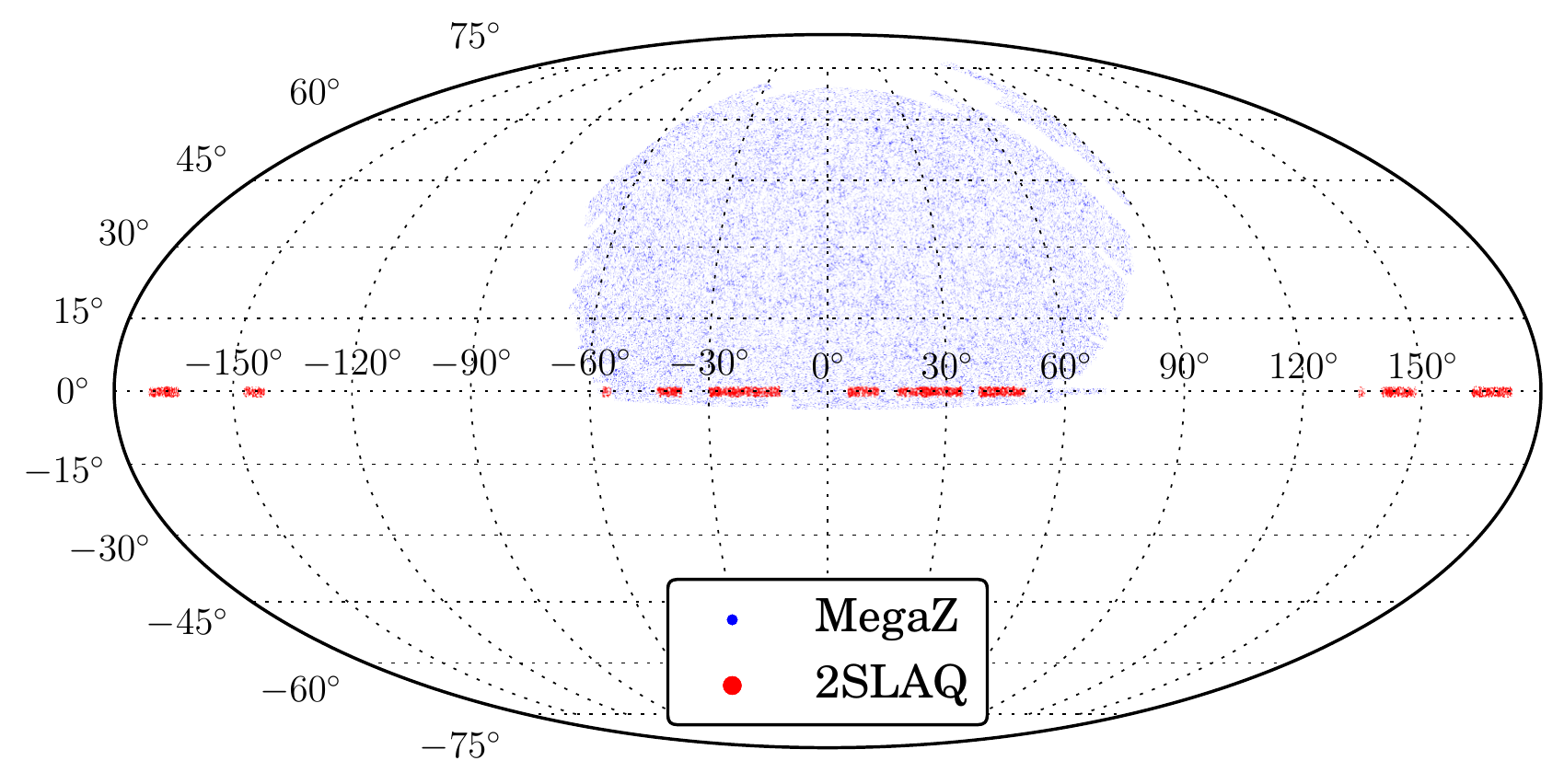}
\caption{Mega-Z and 2SLAQ maps in Mollweide projection plotted in blue and red respectively. For the sake of clarity only a hundred thousand galaxies randomly selected from Mega-Z and five thousand from 2SLAQ have been plotted. The Mega-Z sample covers a total of 7750~deg$^2$, while 2SLAQ covers 180~deg$^2$. The 2SLAQ area is divided into several fields inside a 2$^\circ$-wide strip that extends along the celestial equator.} 
\label{scatter_map}
\end{figure*}

The Mega-Z LRG DR7\footnote{An ASCII version of the Mega-Z LRG DR7 catalog can be found at \url{http://zuserver2.star.ucl.ac.uk/~sat/Mega-Z/Mega-ZDR7.tar.gz.}} catalog \citep{Collister:2006qg}  includes $\sim$1.4 million Luminous Red Galaxies from the SDSS Data Release 7 in the redshift range $0.4< z < 0.7$, with limiting magnitude $i_{AB}<20$. It covers an area of $\sim$7750~deg$^2$ of the sky that is displayed in Fig.~\ref{scatter_map}. This is the sample in which we will investigate the effect of the photo-z quality cuts on the observed galaxy clustering. 

In order to calibrate the photometric redshifts of the Mega-Z galaxies, a representative galaxy sample with known redshifts is needed.  Fortunately, such a sample exists: the 2dF-SDSS LRG and Quasar\footnote{The whole 2SLAQ data can be downloaded from \url{http://www.2slaq.info/query/2slaq_LRG_webcat_hdr.txt.}} (2SLAQ) catalog \citep{Cannon:2006qh}  was obtained using the same selection criteria as the Mega-Z catalog and includes $\sim$13100 LRGs with spectroscopic redshifts. Its sky coverage of only 180~deg$^2$ can be seen in Fig.~\ref{scatter_map} in red. The galaxies are located on a strip of 2~deg along the celestial equator, the area subtended by the 2dF spectrograph. Only non-repeated objects ($ind=1$) with high spectroscopic redshift confidence level ($hqs\ge3$) are used in this analysis. 

\begin{figure*}
\centering
\includegraphics[type=pdf,ext=.pdf,read=.pdf, width=150mm]{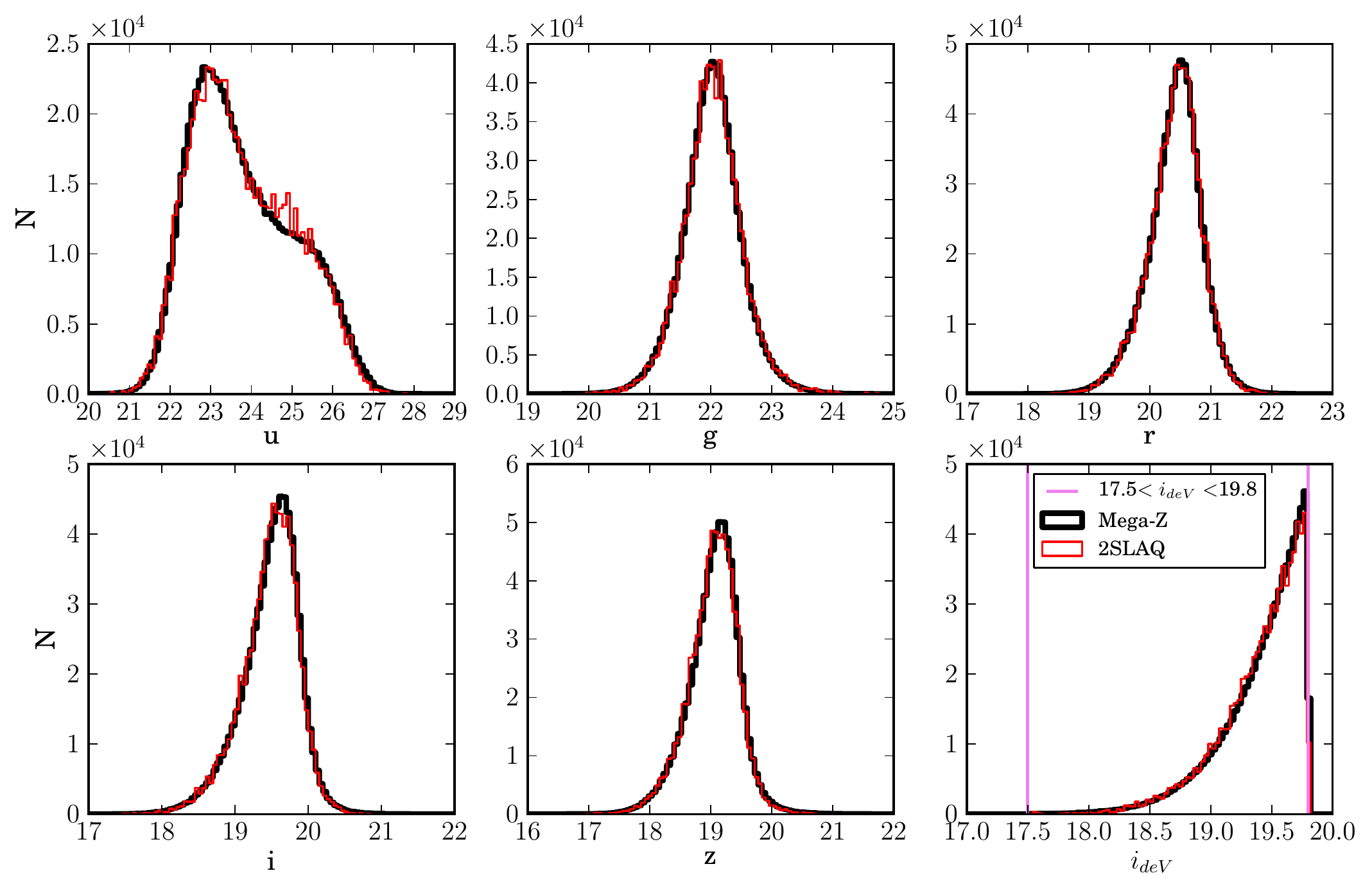}
\caption{From top-left to bottom-right, the model magnitude distributions in the $ugriz$ bands for 2SLAQ in red and Mega-Z in black, normalized to each other. The last plot corresponds to the $i$ band de Vaucouleurs magnitude. All magnitudes are corrected for galactic extinction. The purple lines in the last plot show the magnitude cut 
in~(\ref{mag_cut}), which is the nominal limiting magnitude for the 2SLAQ sample. The agreement is excellent, except in the $u$ band, where the low signal-to-noise produces a small disagreement around magnitude $\sim$25.}
\label{Nm_megaz}
\end{figure*}

The selection criteria in both catalogs consist of a magnitude cut and several color cuts. All magnitudes have been corrected for galactic extinction;  we use model magnitudes for the color cuts and to compute the photometric redshifts (section~\ref{sec:photoz}). The magnitude cut 
\begin{equation}
17.5<i_{deV}<19.8
\label{mag_cut}
\end{equation}
is motivated by the limiting magnitude of the 2dF spectrograph and to ensure completeness of the 2SLAQ catalog. While the Mega-Z catalog is complete up to magnitude $i_{deV}=20$, the 2SLAQ completeness drops off sharply beyond $i_{deV}=19.8$, so the cut forces us to cut the Mega-Z sample at this limit. It eliminates $\sim$32\% of the Mega-Z galaxies leaving a total of $\sim$950000. The $i_{deV}$ magnitude distribution for both samples is plotted on the bottom-right of Fig.~\ref{Nm_megaz}. The purple lines represent the cut in~(\ref{mag_cut}).

The color cuts applied are: 
\begin{eqnarray}
0.5<g-r&<&3 \label{gr_isol_lrg}\\
r-i&<&2 \label{ri_isol_lrg}\\
c_\parallel \equiv 0.7(g-r)+1.2(r-i-0.18)&>&1.6 \label{later_type}\\
d_\perp \equiv (r-i)-(g-r)/8.0&>&0.55 \label{zp_cut} \, .
\end{eqnarray}
They are used to isolate the LRGs from the rest of galaxies. In particular, (\ref{later_type}) separates later-type galaxies from LRGs, and (\ref{zp_cut}) acts as an implicit photo-z cut of $z\gtrsim0.45$, as we will see in the next section.
In~\citet{Collister:2006qg} the $d_\perp$ cut is set to 0.5 for the Mega-Z catalog, but, once again, the 2SLAQ completeness within $0.5<d_\perp<0.55$ is very poor: once all other cuts are applied, they represent 3.6\% of the galaxies, instead of 27\% in Mega-Z. Therefore, we choose $d_\perp > 0.55$.

\begin{figure*}
\centering
\includegraphics[type=pdf,ext=.pdf,read=.pdf, width=130mm]{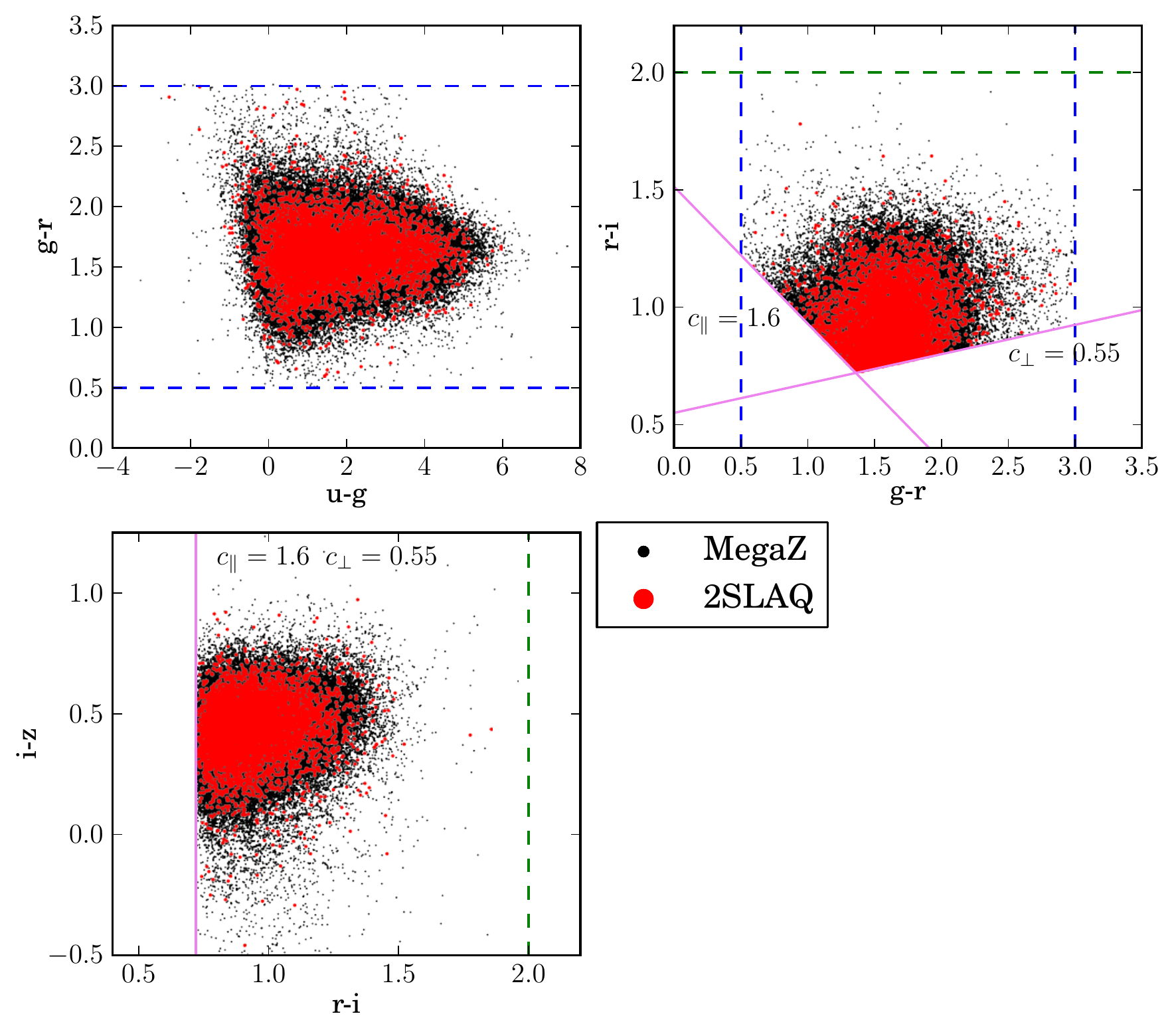}
\caption{Color-color diagrams for both the Mega-Z catalog in black and the 2SLAQ catalog in red. For the sake of clarity, we have only plot a hundred-thousand galaxies for Mega-Z and five thousand for 2SLAQ. We can see that 2SLAQ covers the same color area as Mega-Z, so that we can conclude that it is a good representative spectroscopic sample. The blue and green dashed lines show the color cuts in (\ref{gr_isol_lrg}) and (\ref{ri_isol_lrg}) respectively, while the purple solid lines show the cuts in (\ref{later_type}) and (\ref{zp_cut}), used to select LRGs and high-z galaxies, respectively. The two last cuts shown in the top-right plot translate into an implicit cut of $r-i>0.72$ in the bottom plot.}
\label{colors_megaz}
\end{figure*}

These magnitude and color cuts leave a total of 749152 objects in the Mega-Z catalog and 11810 in the 2SLAQ catalog. In Figs.~\ref{Nm_megaz} and \ref{colors_megaz}, we plot the model magnitude distributions and the color-color scatters, respectively, for all the $ugriz$ bands, after applying all cuts. 2SLAQ is shown in red and Mega-Z in black. Solid and dashed lines represent the cuts. The 2SLAQ magnitude distributions have been normalized up to the total amount of galaxies in Mega-Z. The agreement between both catalogs is excellent, so that we can conclude that 2SLAQ is a representative spectroscopic sample of Mega-Z.

Additionally, some extra cuts have been applied in the Mega-Z catalog to reduce the star contamination:
\begin{eqnarray}
i_{psf}-i_{model}&>&0.2\times(21.0-i_{deV})\\
i\text{-band de Vaucouleurs radius}&>&0.2\\
\delta_{sg} &>& 0.2
\end{eqnarray}
As explained in~\citet{Collister:2006qg}, the first two cuts separate galaxies from stars leaving a residual $\sim$5\% contamination of M-type stars, which cannot be trivially separated either using $gri$ colors or through cuts on the subtended angular diameter. Because of this, the last cut is applied. First, the photo-z neural network ANNz~\citep{Collister:2003cz} is trained on the 2SLAQ catalog that contains reliable information about whether objects are stars or galaxies. The trained network is then run on the whole Mega-Z catalog to compute the probability $\delta_{sg}$ that objects be galaxies. Removing all objects with probability below 0.2 reduces the stellar contamination from 5\% to 2\%~\citep{Collister:2006qg}. In 2SLAQ, we can remove stars by simply getting rid of all those objects with redshift less than 0.01. 

\section{Photometric redshifts}
\label{sec:photoz}

We will perform the galaxy clustering study in several photometric redshift (photo-z) bins. Therefore, we will need to estimate the photo-z of each galaxy in the Mega-Z sample. Furthermore, the theoretical predictions for the clustering need the true-redshift distribution of the galaxies in each photo-z bin $i$, $N_i(z)$, which we can obtain from the 2SLAQ sample. 
For this purpose, we need to compute photometric redshifts of the 2SLAQ galaxies, split them into several photo-z bins, and, finally, recover the spectroscopic redshift distribution in each bin. Additionally, we will study the photo-z performance using 2SLAQ, and apply photo-z quality cuts to improve it. The impact of these cuts on $N_i(z)$ will also be studied.

We use the Bayesian Photometric Redshifts\footnote{\texttt{BPZ} can be found at \url{http://www.its.caltech.edu/~coe/BPZ/}.} (BPZ) template-fitting code described in~\citet{Benitez:1998br} to compute the photometric redshift of galaxies in both catalogs. It uses Bayesian statistics to produce a posterior probability density function $p(z|m_i)$ that a galaxy is at redshift $z$ when its magnitudes are $m_i$:
\begin{equation}
p(z|m_i) \propto \sum_t L(m_i|z,t) \, \Pi(z,t \mid m_i)  \, ,
\label{pz}
\end{equation} 
where $L(m_i|z,t)$ is the likelihood that the galaxy has magnitudes $m_i$, if its redshift is $z$ and its spectral type $t$, and $\Pi(z,t \mid m_i)$ is the prior probability that the galaxy has redshift $z$ and spectral type $t$. Finally, the photometric redshift $z(phot)$ of the galaxy will be taken as the position of the maximum of $p(z|m_i)$.

\begin{figure}
\centering
\includegraphics[type=pdf,ext=.pdf,read=.pdf, width=80mm]{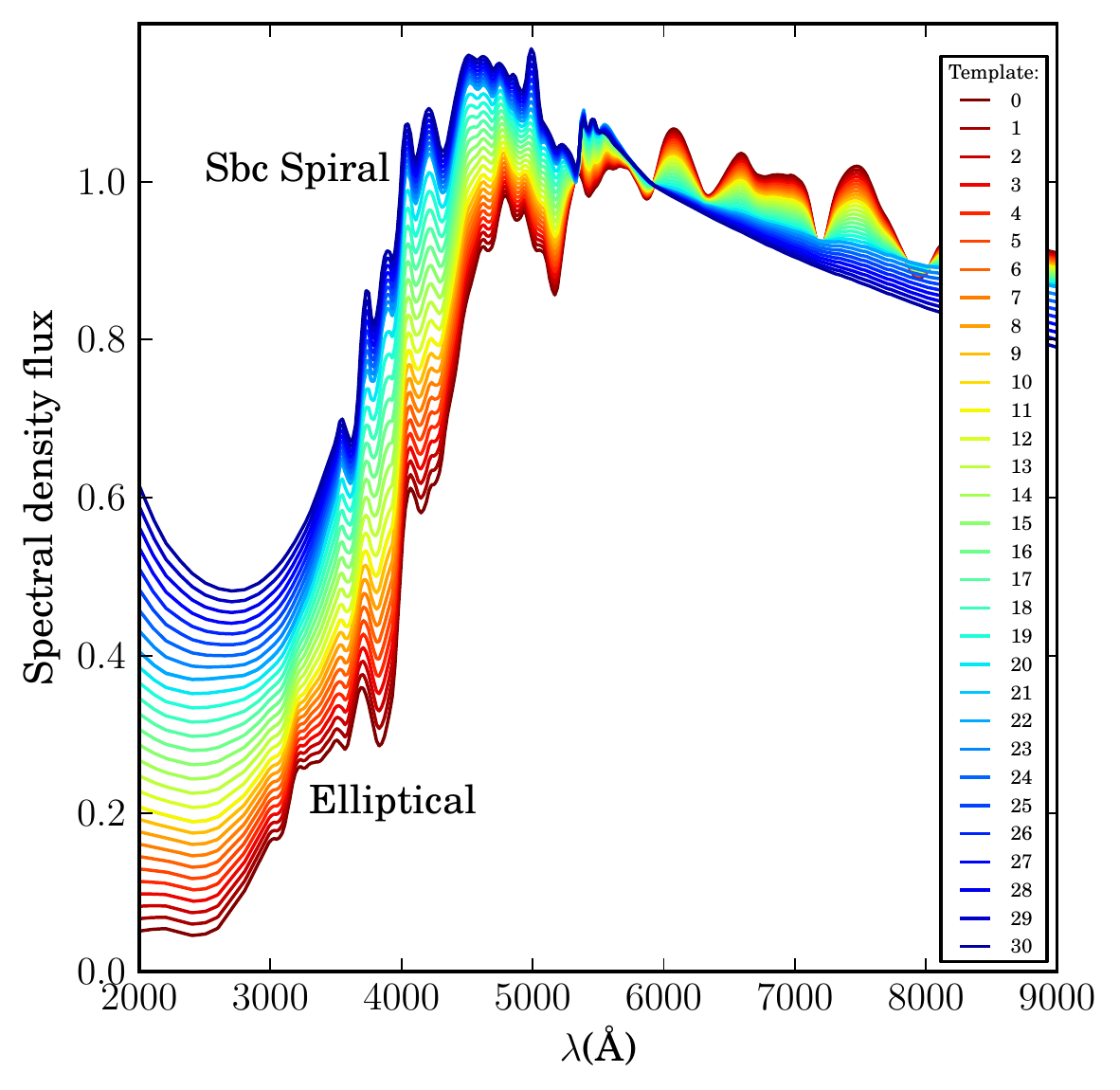}
\caption{The spectral galaxy templates used in the determination of the 2SLAQ and Mega-Z photo-zs. There are a total of 31 templates that range from elliptical galaxies, the vast majority in both samples, to Sbc spiral galaxies.}
\label{2slaq_templates}
\end{figure}

Each spectral type $t$ can be represented by a galaxy template. BPZ includes its own template library, but we prefer to use the new \texttt{CWW} library from \texttt{LePhare}\footnote{The new \texttt{CWW} library can be found in the folder \tt{/lephare\_dev/sed/GAL/CE\_NEW/} of the \texttt{LePhare} package at \url{http://www.cfht.hawaii.edu/~arnouts/LEPHARE/DOWNLOAD/lephare\_dev\_v2.2.tar.gz}.}, another template-based photo-z code described in \citet{Arnouts:1999bb,Ilbert:2006dp}. Both libraries are based on \citet{Coleman:1980,Kinney:1996zz}, but \texttt{BPZ} contains only 8 templates compared to 66 in~\texttt{LePhare}. The large number of templates allows us to focus on the LRG templates, which correspond to the genuine galaxy type of our catalogs. In particular, we select four: \texttt{Ell\_01}, \texttt{Ell\_09}, \texttt{Ell\_19} and \texttt{Sbc\_06}, and then we create nine interpolated templates between consequtive templates, giving a total of 31 templates, shown in Fig.~\ref{2slaq_templates}.

Template-based photo-z codes require the knowledge of the filter band-passes of the survey instrument in order to compute the predicted photometry that will be compared with the observations to produce the likelihood $L(m_i|z,t)$. The SDSS instrument carries five broad-band filters, \textit{ugriz}, described in \citet{Fukugita:1996qt}, whose throughputs are obtained from~\url{http://home.fnal.gov/~annis/astrophys/filters/filters.new}.

A crucial point of \texttt{BPZ} is the prior probability $\Pi(z, t \mid m)$ that helps improve the photo-z performance. \citet{Benitez:1998br} proposes the following empirical function:
\begin{equation}
\Pi(z, t \mid m) \propto f_t e^{-k_t(m-m_0)} \cdot z^{\alpha_t}\exp \left\lbrace -\left[ {z \over z_{mt}(m)} \right]^{\alpha_t} \right\rbrace,
\label{prior}
\end{equation}
where $z_{mt}(m) = z_{0t} + k_{mt}(m-m_0)$. Every spectral type $t$ has associated a set of five parameters $\lbrace f,k,\alpha,z_0,k_{m} \rbrace$ that determine the shape of the prior. These parameters are determined from the spectroscopic data themselves. In principle, we could assign one prior $\Pi(z, t \mid m)$ to each one of the 31 templates in Fig.~\ref{2slaq_templates}, but this would give a total of $31\times5=155$ parameters, too many for the spectroscopic data sample available. Instead, we split the 31 templates in two groups: $t=1$, with the 10 first templates that make up the group of pure elliptical galaxies, and $t=2$ with the rest. Then, running \texttt{BPZ} on 2SLAQ a first time, without priors, we find the group each galaxy belongs to. Fitting~(\ref{prior}) to this output, together with the spectroscopic redshifts and the observed magnitudes in one band (we choose the $i$ band), we find the values of the prior parameters. Results are given in Table~\ref{tab:prior}.
\begin{table}
\centering
\begin{tabular}{|c||ccccc|}
\hline
$t$ & $f$ & $k$ & $\alpha$ & $z_0$ & $k_{m}$ \\ \hline
1 & 0.72 & 0.0 & 8.679 & 0.477 & 0.078\\
2 & 0.14 & 0.0 & 7.155 & 0.488 & 0.064\\
\hline
\end{tabular}
\caption{The values of the prior parameters of (\ref{prior}) for the 2SLAQ galaxies.}
\label{tab:prior}
\end{table}

The $k$ parameters are related to the migration of galaxies from one spectral type to another at different magnitudes. The fit gives values of $k$ very close to 0, so we impose explicitly not having type migration by setting them exactly to 0. $f$ gives the fraction of galaxies of each type at magnitude $m_0$, which we choose to be $m_0=18.5$. Since $k=0$, we find that the 72\% of the galaxies belong to the spectral type group 1 independently of the magnitude, confirming that most of the galaxies are purely elliptical.

We define galaxies with catastrophic redshift determinations as those with $|\Delta z| \equiv |z(phot) - z(spec)| > 1$, where $z(spec)$ is the spectroscopic redshift and $z(phot)$ the photo-z. With the help of the prior, we are able to remove all these catastrophic redshift determinations, which account for $\sim$4.2\% of the 2SLAQ sample. They are typically galaxies with degeneracies in their color space, which cause confusions in the template fit and result in a photo-z much larger than the real redshift. Defining the photo-z precision $\sigma_z$ as half of the symmetric interval that encloses the 68\% of the $\Delta z$ distribution area around the maximum, we also find that its value for the non-catastrophic determinations improves by a factor 1.7, down to $\sigma_z \sim0.042$, in agreement with \citet{Padmanabhan:2004ic,Collister:2006qg,Thomas:2011cu}.

We also apply a cut on the quality of the photometry, consisting of not using any band, for each galaxy, with magnitude error $>$0.5. This cut mostly removes the information from the \textit{u} band for many galaxies, since the signal-to-noise tends to be lower in this band. The overall precision improves slightly to $\sigma_z \sim0.041$.

\begin{figure}
\centering
\includegraphics[type=pdf,ext=.pdf,read=.pdf, width=80mm]{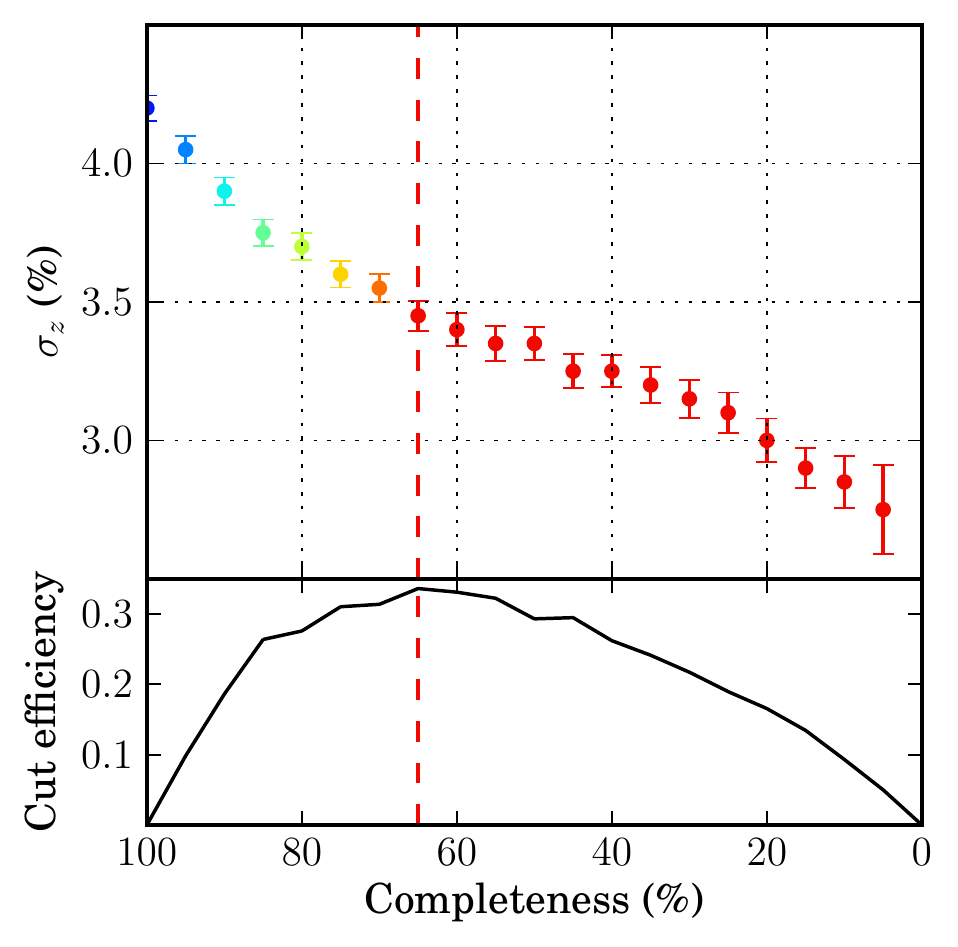}
\caption{Top plot: the 2SLAQ photo-z precision $\sigma_{z}$ for different photo-z quality cuts resulting in the completeness shown on the $x$ axis. The error bars are computed using bootstrap~\citep{efron79}. The color scale labels the different photo-z quality cuts, here and also in Figs.~\ref{2slaq_pz_results} and \ref{Nz_bins}. The nominal precision, without any \textit{odds} cut, is 0.042. It can be improved by a factor 1.5 when the most aggressive cut, which leaves only 5\% of the galaxies, is applied. However, on the bottom plot, we see that the most efficient cut (defined in (\ref{cut_efficiency})) is at 65\% completeness, where by removing 35\% of the galaxies we achieve 50\% of the improvement, with $\sigma_z\sim0.035$. Only cuts with completeness $\geq65\%$ are considered in the following.}
\label{sigvseff}
\end{figure}

Photo-z codes, besides returning the best estimate for the redshift, typically also return an indicator of the photo-z quality. It can be simply an estimation of the error on $z(phot)$, or something more complex, but the aim is the same. In \texttt{BPZ}, this indicator is called \textit{odds}, and, it is defined as
\begin{equation}
odds = \int^{z(phot)+\delta z}_{z(phot)-\delta z}p(z|m_i)dz \, ,
\label{odds}
\end{equation} 
where $\delta z$ determines the redshift interval where the integral is computed. \textit{Odds} can range from 0 to 1, and the closer to 1, the more reliable is the photo-z determination, since $p(z|m_i)$ becomes sharper and most of its area is enclosed within $z(phot)\pm \delta z$. In our case, we choose $\delta z = 0.03$, which is close to the photo-z precision in 2SLAQ and Mega-Z. A bad choice of $\delta z$ could lead to the accumulation of all \textit{odds} close to either 0 or 1. Since \textit{odds} are a proxy for the photo-z quality, we should expect a correlation between the \textit{odds} and $\Delta z$, in the sense that higher \textit{odds} should correspond to lower $|\Delta z|$. In Fig.~\ref{sigvseff}, we show $\sigma_z$ for subsets of the Mega-Z sample with increasingly higher cuts on the \textit{odds} parameter. In fact, the exact \textit{odds} values are quite arbitrary, since they depend on the size of $\delta z$. Therefore, we have translated these \textit{odds} cuts into the fraction of the galaxy sample remaining after a certain cut has been applied. The abcissa in Fig.~\ref{sigvseff} corresponds to this completeness for increasingly tighter \textit{odds} cuts. 

We see that, by removing the galaxies with low \textit{odds} in steps of 5\% in completeness, we are able to reduce the photo-z dispersion from $\sigma_z \sim$ 0.042 to $\sim$0.028, a factor of 1.5. Obviously, the best accuracy is obtained when the completeness is close to 0\%, but this is very inefficient. Defining the efficiency of the cut as:
\begin{equation}
\text{Cut Efficiency } (x) = x \left[ {\sigma_z (100\%) - \sigma_z(x) \over \sigma_z (100\%) - \sigma_z(0\%)} \right],
\label{cut_efficiency}
\end{equation}
where $x$ is the completeness of the catalog after the cut, we find that the most efficient photo-z quality cut is at 65\% of completeness, where $\sigma_z \sim 0.035$, as shown in the bottom plot in Fig.~\ref{sigvseff}. Since we cannot compute $\sigma_z(0\%)$ for lack of galaxies, we use instead in~(\ref{cut_efficiency})
the value at 5\% completeness. From now on, we will refer to photo-z quality cuts as all those in Fig.~\ref{sigvseff} that lead to a completeness between 100\% and 65\%, and which are labeled in different colors.

\begin{figure*}
\centering
\includegraphics[type=pdf,ext=.pdf,read=.pdf, width=135mm]{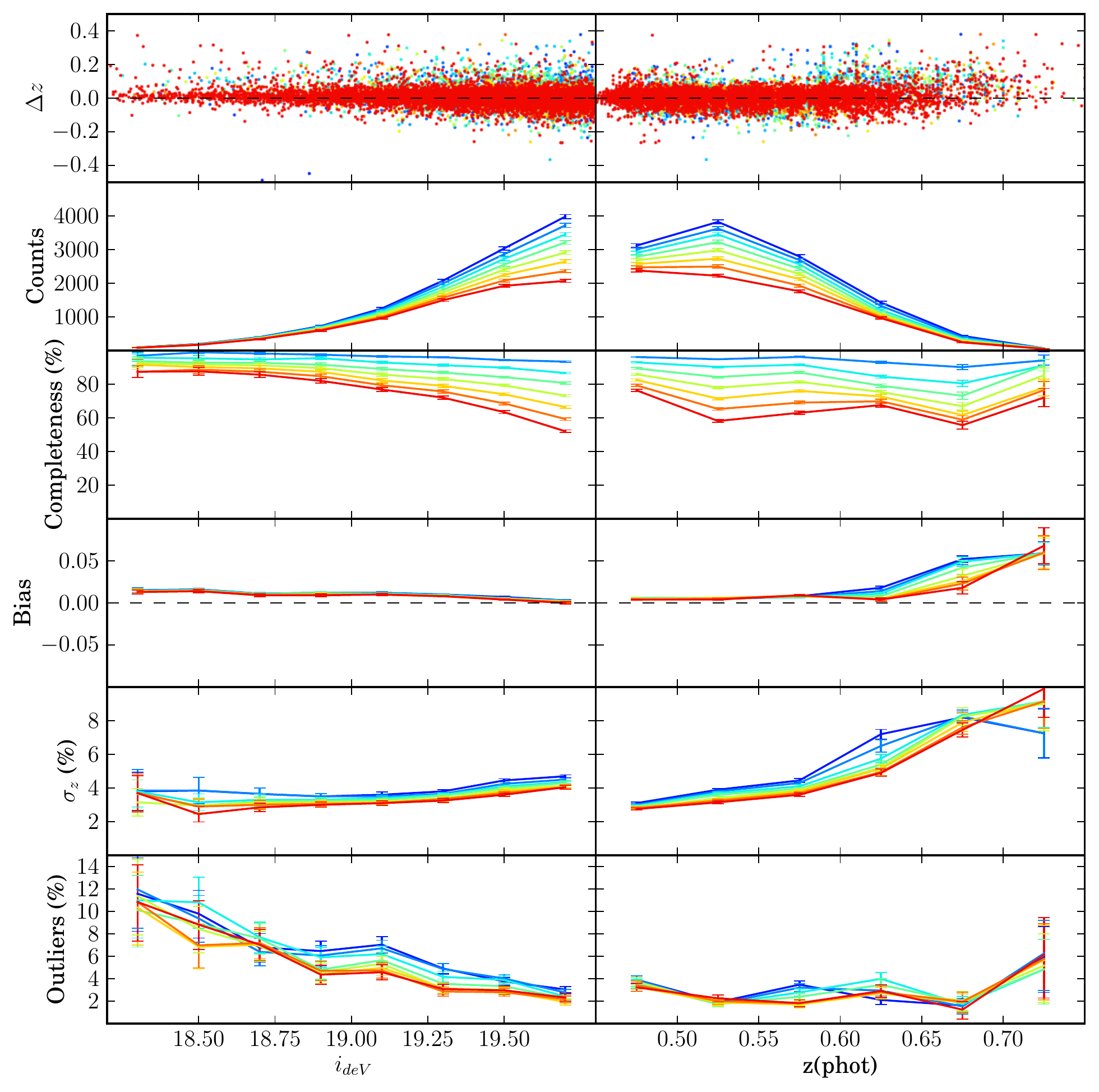}
\caption{Statistics showing the 2SLAQ photo-z performance. In the first row we show the scatter of $\Delta z \equiv z(phot) - z(spec)$ with respect to the $i_{deV}$ magnitude (left) and $z(phot)$ (right). The scatter has been binned along these two variables and some statistical estimators have been computed in each bin. In descending order of rows, we show the galaxy population (in counts), the completeness, the bias (median), the photo-z precision $\sigma_z$ and the 3$\sigma$ outlier fraction. The color degradation from red to blue is the same as in Fig.~\ref{sigvseff} and labels different photo-z quality cuts.}
\label{2slaq_pz_results}
\end{figure*}

An exhaustive analysis of the photo-z results is shown in Fig.~\ref{2slaq_pz_results}. It consists of a series of plots where different statistical properties of $\Delta z$, in the rows, are shown as a function of two different variables, in the columns: $i_{deV}$ magnitude on the left, and the photo-z estimation, $z(phot)$ on the right. In the first row, we can see the $\Delta z$ scatter from which the rest of plots are derived. As in Fig.~\ref{sigvseff}, the color progression of the curves from blue to red corresponds to the different photo-z quality cuts with completeness going from 100\% to 65\%, the most efficient cut, in steps of 5\%. The number of galaxies, in the second row, grows for increasing magnitudes, but drops for increasing redshifts. The completeness, in the third row, drops at high magnitudes,
especially when quality cuts become harder. It is quite constant along redshift. The bias (median), in the forth row, shows a general offset of $\sim$ 0.02, so that photo-zs are in general $\sim$4\% larger than the actual redshift. The photo-z precision $\sigma_z$, in the fifth row, degrades slightly for fainter galaxies, while it does by a factor of almost 3 for high-z galaxies. As in Fig.~\ref{sigvseff}, the harder the photo-z quality cut, the better precision we get. Finally, the last estimator is the outlier fraction, defined as the fraction of galaxies with $|\Delta z|$ above three times $\sigma_z$. It decreases from 10\% to 3\% for increasing magnitudes, while it keeps constant around 3\% along redshift. The photo-z quality cuts help reduce it at some magnitudes and redshifts. 

\begin{figure}
\centering
\includegraphics[type=pdf,ext=.pdf,read=.pdf, width=85mm]{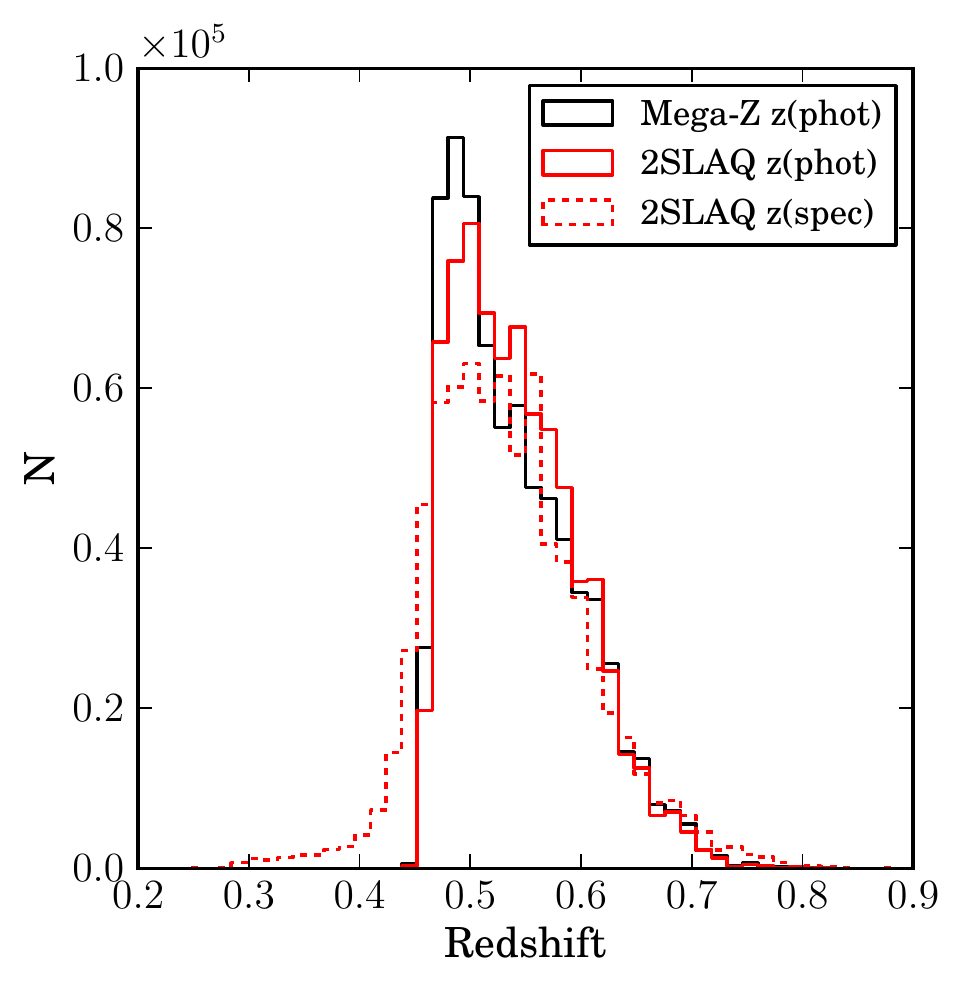}
\caption{The photo-z distribution of Mega-Z (black), and the spectroscopic (red dashed) and photometric (red solid) redshift distributions of 2SLAQ. The 2SLAQ distributions have been normalized to the Mega-Z number of galaxies for comparison.}
\label{Nz_megaz}
\end{figure}

In Fig.~\ref{Nz_megaz}, we have compared the 2SLAQ photo-z distribution with the spectroscopic redshift distribution. Both distributions are clearly different at low redshift. While the photo-z distribution rises very sharply from $z\sim0.45$, reaching the maximum immediately, the spectroscopic distribution rises much more gradually from $z\sim0.25$. This is because the color cut in (\ref{zp_cut}) acts as a photo-z cut at $z(phot)\gtrsim0.45$. Figure~\ref{Nz_megaz} also contains the photo-z distribution of the Mega-Z galaxies. It closely resembles the photo-z distribution in 2SLAQ, but they are not as similar as the magnitude distributions in Fig.~\ref{Nm_megaz}. 

\begin{figure}
\centering
\includegraphics[width=80mm]{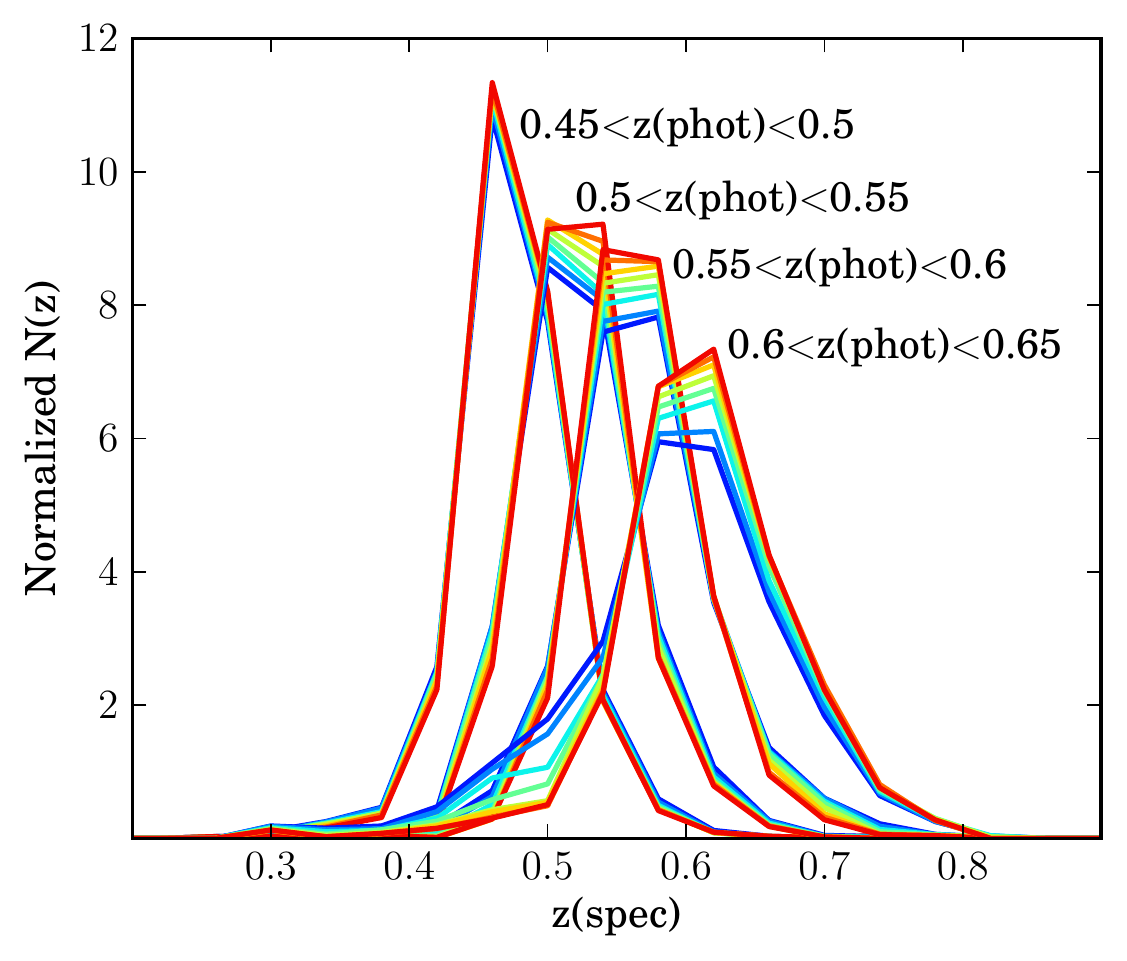}
\caption{The spectroscopic redshift distributions in the 2SLAQ catalog for the four photo-z bins in which we will measure galaxy clustering in section~\ref{sec:clustering}. They are all normalized to the same area under the curves. Different colors label different photo-z quality cuts, as in Figs.~\ref{sigvseff} and \ref{2slaq_pz_results}.}
\label{Nz_bins}
\end{figure}

Finally, we split the 2SLAQ and Mega-Z catalogs into four photo-z bins of equal width 0.05. The width has been chosen to roughly match the photo-z precision of the catalogs. We want to know the actual redshift distributions inside each of these photo-z bins in order to make the predictions for clustering in the next section. For this purpose, we use the spectroscopic information in 2SLAQ. In Fig.~\ref{Nz_bins} we show the spectroscopic redshift distributions of these four photo-z bins in 2SLAQ at the different photo-z quality cuts of Fig.~\ref{sigvseff}, using the same color labeling. All the distributions have been normalized in order to compare them. As expected, the distributions become wider in the higher photo-z bins. This is in agreement with the increasing $\sigma_z$ with redshift seen in Fig.~\ref{2slaq_pz_results}. On the other hand, photo-z quality cuts tend to reduce the width of the distributions. For instance, the left tail of the last bin, at $0.6<z(phot)<0.65$, is drastically reduced when the most efficient photo-z quality cut is applied. 

\section{Galaxy clustering and the effect of the photo-z quality cuts}
\label{sec:clustering}

We will now compute the angular galaxy correlations in the four photo-z bins of Fig.~\ref{Nz_bins}, before and after applying the different photo-z quality cuts of Fig.~\ref{sigvseff}. We want to see and characterize the impact of these cuts on clustering.

For this purpose, we use the Hierarchical Equal Area isoLatitude Pixelization (Healpix) framework~\citep{Gorski:2004by}, developed for CMB data analysis. It provides pixelations of the sphere with pixels of equal area, with their centers forming a ring of equal latitude. The resolution of the grid is expressed by the parameter $N_{side}$, which defines the number of divisions along the side of a base-resolution pixel that is needed to reach a desired high-resolution partition. The total number of pixels in the sphere is given by $N_{pix} = 12 N_{side}^2$. We will use $N_{side} = 256$ for our maps, which divides the sphere into 786432 pixels. 

\begin{figure}
\centering
\includegraphics[width=80mm]{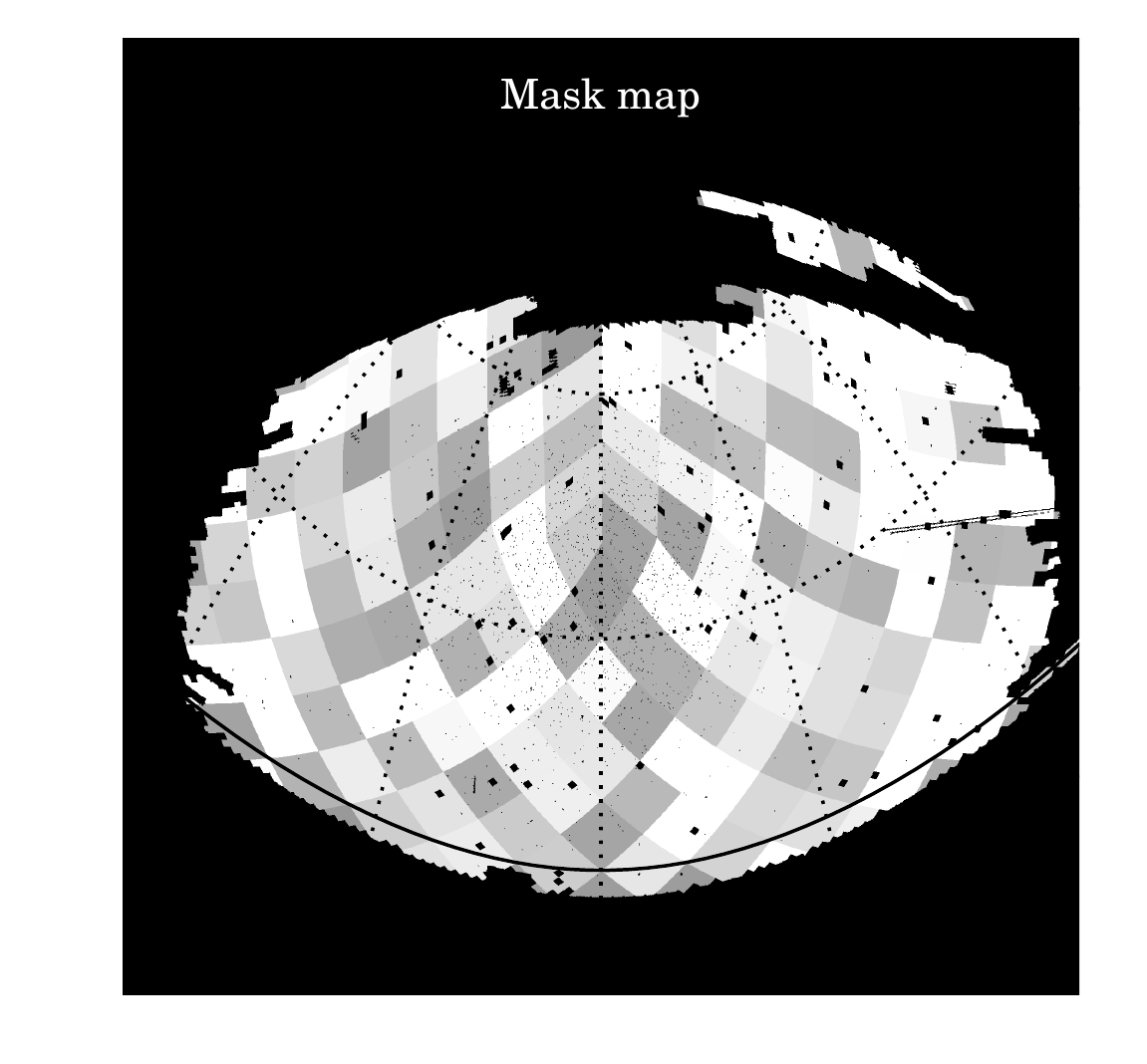}
\caption{The Mega-Z DR7 mask in Healpix of $N_{side}$=512. It is obtained in two steps. First, using a low resolution Healpix map of $N_{side}=64$, we reject all those pixels with less than 65 galaxies per pixel to get rid of the underpopulated areas (black patches) and obtain the overall geometry of the Mega-Z footprint. Second, we repeat the process with a star map to reject poor data quality regions of even smaller size (black dots and the long thin horizontal stripe on the right side). Different gray levels display the 174 jackknife zones used in~(\ref{covariance}) to compute the covariance of the angular correlations between different scales. They are low resolution pixels of a Healpix map with $N_{side} = 8$.}
\label{mask_map}
\end{figure}

Unlike some CMB maps, the galaxy maps do not usually cover the whole sky, so we need to define a mask for the observed area. Moreover, some surveyed regions, for some technical reasons, sometimes become deprived of galaxies. This may cause systematic distortions on the measured galaxy clustering if they are not removed from the mask. We construct the mask in several steps:
\begin{itemize}
\item First, the geometry of the Mega-Z sample is inferred by populating a low resolution Healpix map of $N_{side}=64$ with all the Mega-Z galaxies. All pixels with less than 65 galaxies per pixel are rejected.  
The threshold is chosen in order to cut out a long, low amplitude tail in the distribution of number of objects per pixel.
This low resolution mask has the outline of the SDSS footprint, and also throws away underpopulated sky areas, which can be seen in~Fig.~\ref{mask_map} as small black patches inside the Mega-Z area. These regions include data with poor quality or completeness. Some of the patches will be removed inappropriately, since they may be underpopulated due to normal fluctuations in the number counts of galaxies. We have tested that changes in the cut value used induce differences on the measured correlations that are small compared to the size of the correlation errors. 
\item Second, there are some areas in the SDSS footprint that have poor quality data, but are smaller than a Healpix pixel with $N_{side}=64$. To eliminate these, we download 50 million stars from the SDSS database\footnote{http://casjobs.sdss.org/.} with magnitude down to 19.6. The star catalog is much more spatially dense than the Mega-Z catalog;  therefore, we can construct a mask of the SDSS footprint using the stars in the same manner, but with better resolution, than with the Mega-Z galaxies.  We construct a map of the stars with $N_{side}$=512, and declare pixels as bad if they have less than 7 stars per pixel.  This throws away bad regions such as the long thin horizontal stripe in the right side of Fig.~\ref{mask_map}. It also throws away some pixels at high galactic latitude (black dots in the center) that may lack stars due to normal fluctuations in star counts.  However, the density of stars should be unrelated to the positions of the Mega-Z galaxies, and therefore throwing away a small area with the lowest stellar density should not bias measurements of the galaxy correlation function. 
\item Finally, we reduce the resolution of the mask to $N_{side}=256$, which is the resolution that we will use in our galaxy maps.
\end{itemize}
More details and justification for computing the mask as described can be found in~\citet{Cabre:2008sz}.

\begin{figure*}
\centering
\begin{tabular}{rl}
\includegraphics[width=170mm]{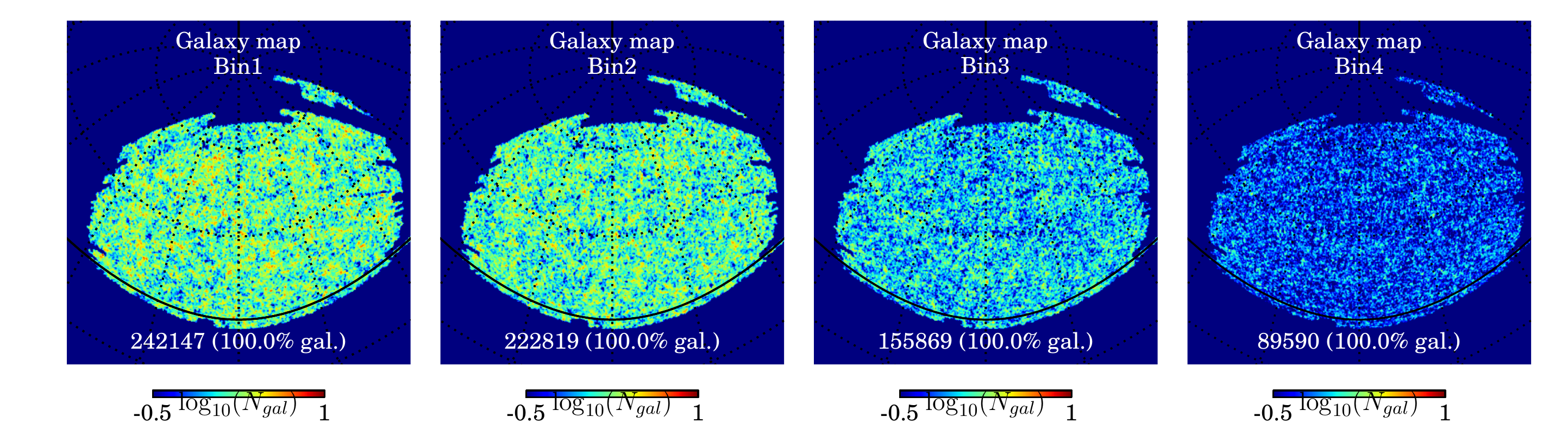} \\
\includegraphics[width=170mm]{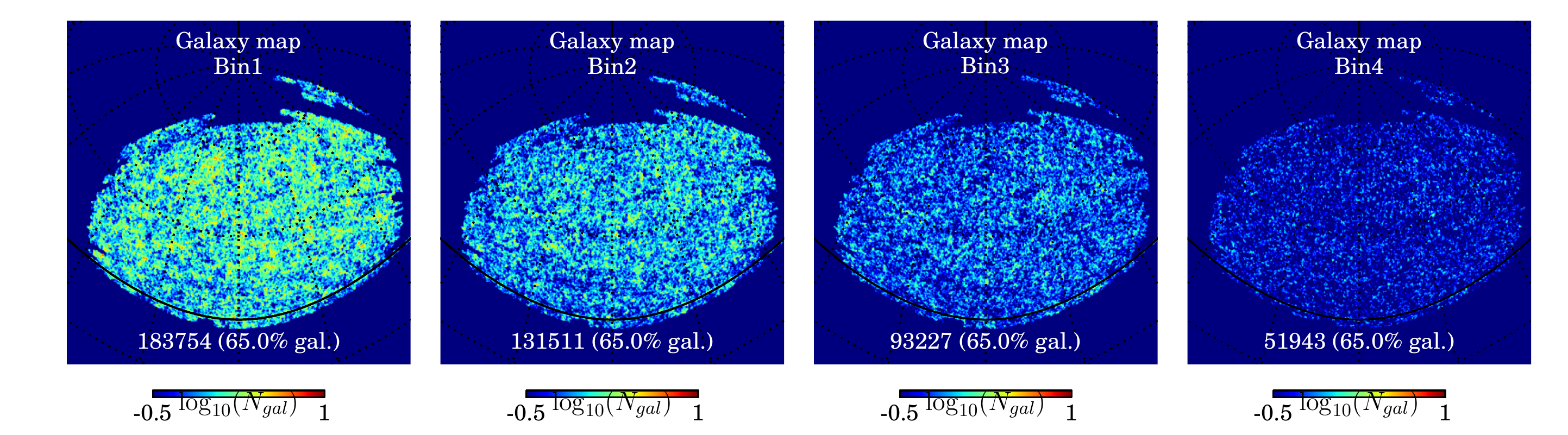} \\
\includegraphics[width=170mm]{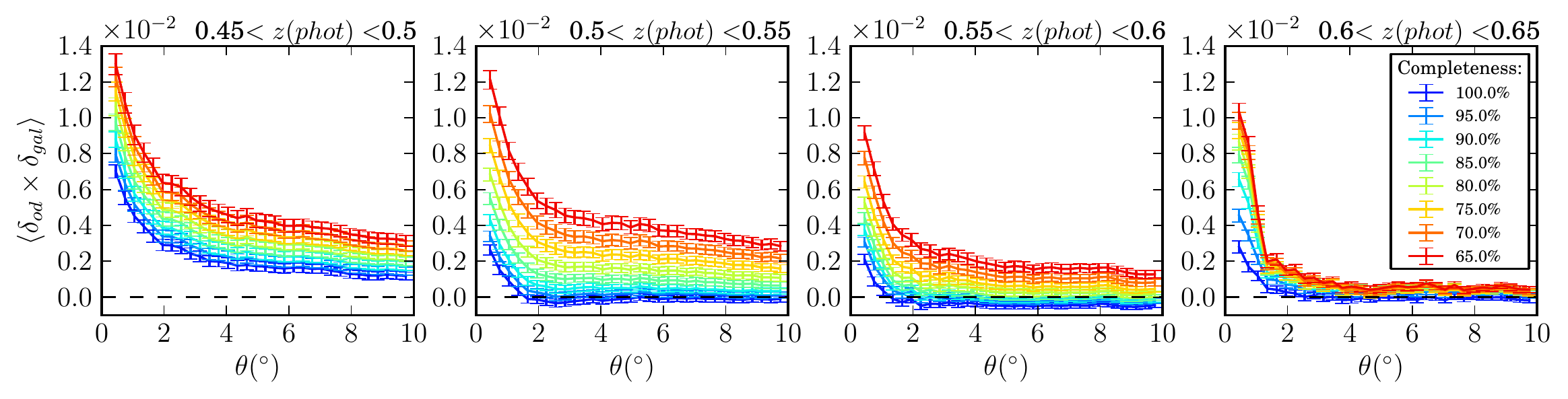}
\end{tabular}
\caption{On the first row, the galaxy maps of the Mega-Z catalog for the four photo-z bins of Fig.~\ref{Nz_bins}. On the second row, the same after applying a photo-z quality cut with 65\% completeness. These are Healpix maps of $N_{side}=256$. The number of galaxies is given in each map. 
On the bottom row, the angular cross correlations of the galaxy maps with the \textit{odds} maps of Fig.~\ref{od_map}, at the different photo-z quality cuts of Fig.~\ref{sigvseff}. Initially, both maps are not cross-correlated at scales $>2^\circ$ (except in the first bin, $0.45<z(phot)<0.5$), but the {\it odds} cut introduce progressively larger cross-correlations between them.}
\label{gal_map}
\end{figure*}

\begin{figure*}
\centering
\includegraphics[width=170mm]{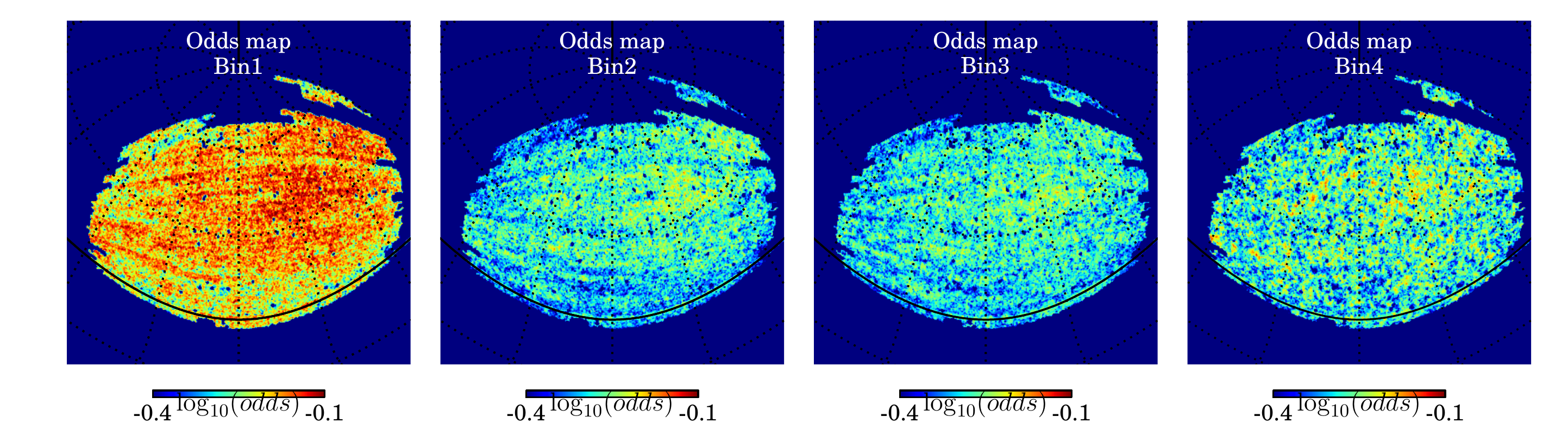} \\
\includegraphics[width=170mm]{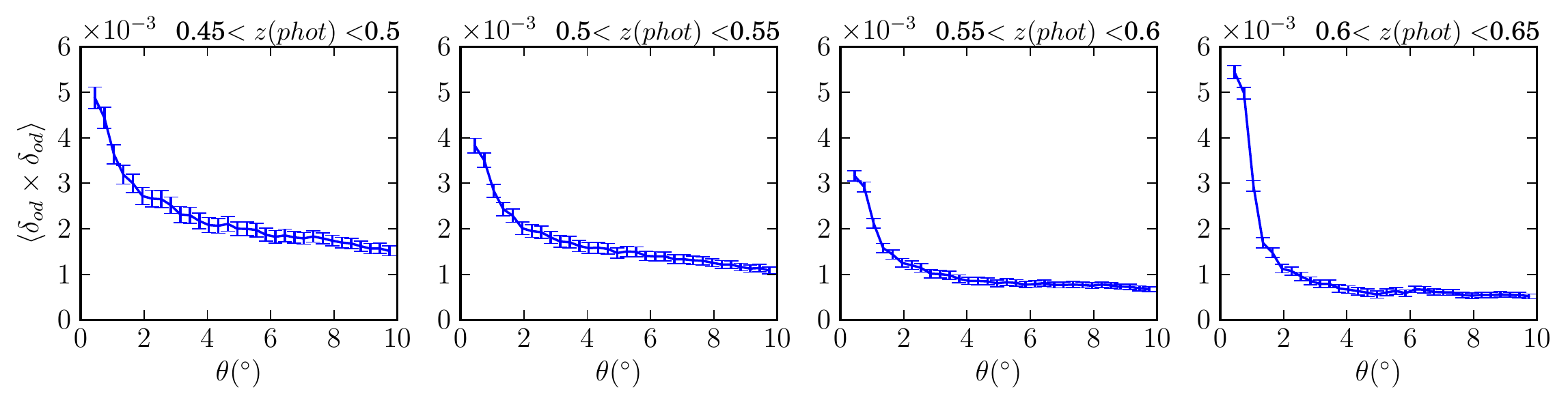}
\caption{On the first row, the \textit{odds} maps of the Mega-Z catalog for the four photo-z bins of Fig.~\ref{Nz_bins}. These are Healpix maps of $N_{side}=256$ where the \textit{odds} values per pixel are computed as the mean \textit{odds} of all the galaxies in each pixel. Redder regions are regions with higher photo-z quality. They are not homogeneously distributed over the mask. 
On the bottom row, the angular auto correlations of the \textit{odds} maps. They are auto correlated at all scales $<10^\circ$, however the strength of these correlations is lower at higher $z(phot)$.}
\label{od_map}
\end{figure*}

Once we have the mask, we create the galaxy maps shown at the top of Fig.~\ref{gal_map} and the \textit{odds} maps shown at the top of Fig.~\ref{od_map}. The galaxy maps are created by counting the number of galaxies that fall in each pixel of the mask, while the \textit{odds} maps are created by averaging the \textit{odds} of these galaxies in the pixel. When no galaxies fall in a pixel, we still need an \textit{odds} value in that pixel, so we take the average value in the neighboring pixels within a circle of 1$^\circ$ radius. At the working resolution, this is a total of 19 pixels, enough so that at least one contains some galaxies, even in the last bin $0.6<z<0.65$, where the average number of galaxies per pixel is $\sim 0.36$ when the 65\% completeness cut is applied. We also create galaxy maps after applying each photo-z quality cut defined in Fig.~\ref{sigvseff}. On the second row of plots in Fig.~\ref{gal_map} we show the galaxy maps after applying the most efficient cut with 65\% completeness. 

The first \textit{odds} map at $0.45<z<0.5$ is clearly redder than the other three, since it contains galaxies with higher photo-z quality. On the contrary, bluer regions mark regions with bad photo-z quality. Note that they are not uniformly distributed on the map. They form a pattern of horizontal strips that cross the entire Mega-Z footprint. This is most noticeable in the bins $0.45<z<0.5$ and $0.5<z<0.55$. Therefore, when we remove low-odds galaxies, we are not taking them uniformly off the map. We can already see this in the galaxy maps of Fig.~\ref{gal_map} when the {\it odds} cut is applied. If we focus on the bin $0.5<z<0.55$, we see that the map shows a strip pattern very similar to the bluer zones of its corresponding {\it odds} map. In Fig.~4 of \citet{Crocce:2011mj} the authors show a map of the mean error on the r magnitude per pixel of a catalog similar to Mega-Z. Besides the regions with clearly bad photometry due to galactic extinction, regions with low photometric quality in the center of the footprint form patterns very similar to those on our odds maps, with horizontal strips crossing the whole Mega-Z footprint. They approximately coincide with the drift scan paths of the SDSS instrument, and, due to observations done at different nights with different photometric quality of the atmosphere, they have resulted in regions of poor photo-z quality on the sky.

Next we want to compute the angular correlations on all these maps. The angular correlations between two Healpix maps, $a$ and $b$, are given by:
\begin{equation}
\omega_{ab}(\theta) \equiv \langle \delta_a \delta_b \rangle (\theta) = {1 \over N_{\theta}} \sum^{N_{\theta}}_{i,j} \delta_{a,i} \delta_{b,j} \, ,
\label{measured_correlations}
\end{equation}
where $\delta_{a,i} = a_i/\bar{a} - 1$ is the fluctuation of the map $a$ at the pixel $i$ with respect to the mean $\bar{a}$, and $N_{\theta}$ is the total number of combinations of pixels $i$ and $j$ separated by an angular distance between $\theta$ and $\theta+\Delta \theta$. Since the typical pixel resolution  when $N_{side}=256$ is $\sim 0.2^\circ$, we choose $\Delta \theta$ to be $0.3^\circ$. We also want to know the covariance of $\omega_{ab}(\theta)$. We can compute it using the \textit{jackknife} technique, also used in a similar study in~\citet{Crocce:2011mj} and explained in detail in~\citet{Cabre:2007rv}. This technique consists of spliting the survey area within the mask into $N_k$ sub-areas. We have used pixels of a low resolution Healpix map of $N_{side}=8$ as the different jackknife sub-areas. We end up with a total of 174 pixels lying on the mask, represented by different gray levels in Fig.~\ref{mask_map}. The correlations will be computed $N_k$ times, each time removing each one of the sub-areas. This will result in the jackknife correlations $\omega_k(\theta)$. Then, the covariance between the correlation functions measured at angles $\theta_n$ and $\theta_m$ will be:
\begin{equation}
\text{Cov}_\omega(\theta_n, \theta_m) = {(N_k-1) \over (N_k)} \sum^{N_k}_{k=1}\left[ \omega_k(\theta_n) - \bar{\omega}(\theta_n) \right]\left[ \omega_k(\theta_m) - \bar{\omega}(\theta_m) \right] \, ,
\label{covariance}
\end{equation}
where $\bar{\omega}(\theta) = \sum_k^{N_k} \omega_k(\theta) / N_k$ is the mean of all the jackknife correlations. Therefore, the diagonal errors of the measured correlation at $\theta$ will be $\sigma_\omega (\theta) = \sqrt{\text{Cov}_\omega(\theta, \theta)}$.

On the bottom row of Fig.~\ref{od_map} we show the resulting auto-correlations of the odds maps. They are not zero at any scale, which confirms that the photo-z quality is not uniformly distributed on the sky. The higher the redshift, the lower the auto correlation. However, the highest auto correlation is reached at scales $<1^\circ$ in the fourth bin, $0.6<z<0.65$. 

On the bottom row of Fig.~\ref{gal_map}, we show the resulting galaxy-odds cross-correlations, the cross-correlations between the maps on the top of this figure and those on Fig.~\ref{od_map}. Different curves of different colors label correlations after the different photo-z quality cuts in Fig.~\ref{sigvseff}. The redder curve corresponds to the most efficient cut with 65\% completeness. Apart from the first bin $0.45<z<0.5$, the general tendency is that, initially, there is little correlation between the odds and the galaxies at scales $>2^\circ$, but once the quality cuts are applied, the cross correlations start growing. The harder the cut, the higher the correlations. However, this growth is less noticiable at higher $z$. The reason is that, the spatial features of the odds maps become imprinted into the galaxy maps as the \textit{odds} cuts are applied. So, if the auto correlations of the odds maps are small, the strength of this imprinting will also be small. This agrees with the behaviour of the odds auto-correlations in Fig.~\ref{od_map}. The lower $z$ bin is unusual in the sense that the galaxy-odds correlations are not initally 0 at any scale. This means that regions with an under or over density of galaxies concide with regions with the best or worst photo-z quality. This will be discussed in more detail in the following sections. Even so, the correlations still grow when the cuts are applied.

Finally, we compute the angular galaxy auto- and cross-correlation between the four photo-z bins at different photo-z quality cuts. We also compare our measured correlations with predictions. We compute the predictions for the angular correlation $\omega_{ab}(\theta)$ as described in~\cite{Crocce:2010qi}:
\begin{equation}
\omega_{ab}^{(theo)} (\theta) = \int dz_1 N_a(z_1) \int dz_2 N_b(z_2) \xi^s(z_1,z_2,\theta) \, ,
\label{corr_prediction}
\end{equation}
where $N_i(z)$ are the selection functions, which in our case are the curves in Fig.~\ref{Nz_bins}, $\xi^s(z_1,z_2,\theta)$ is the redshift space correlation of the pairs of galaxies at redshift $z_1$ and $z_2$ subtending an angle $\theta$ with the observer. We use the non-linear power spectrum~\citep{Smith:2002dz} with $\Lambda$CDM with $\Omega_M$ = 0.25, $\Omega_\Lambda$ = 0.75 and $H_0$ = 70~(km/s)/Mpc, the linear \citet{Kaiser:1984sw} model of redshift space distortions for the correlation function, and a linear bias model with evolution: $b(z) = 1.5 + 0.6(z-0.1)$ \citep{Cabre:2008sz}. We have not included the effect of magnification due to gravitational lensing, which turns out to be negligibly small for this sample. Note that the inclusion of photo-z errors in the predictions is only through the $N_i(z)$. 

\begin{figure*}
\centering
\includegraphics[width=170mm]{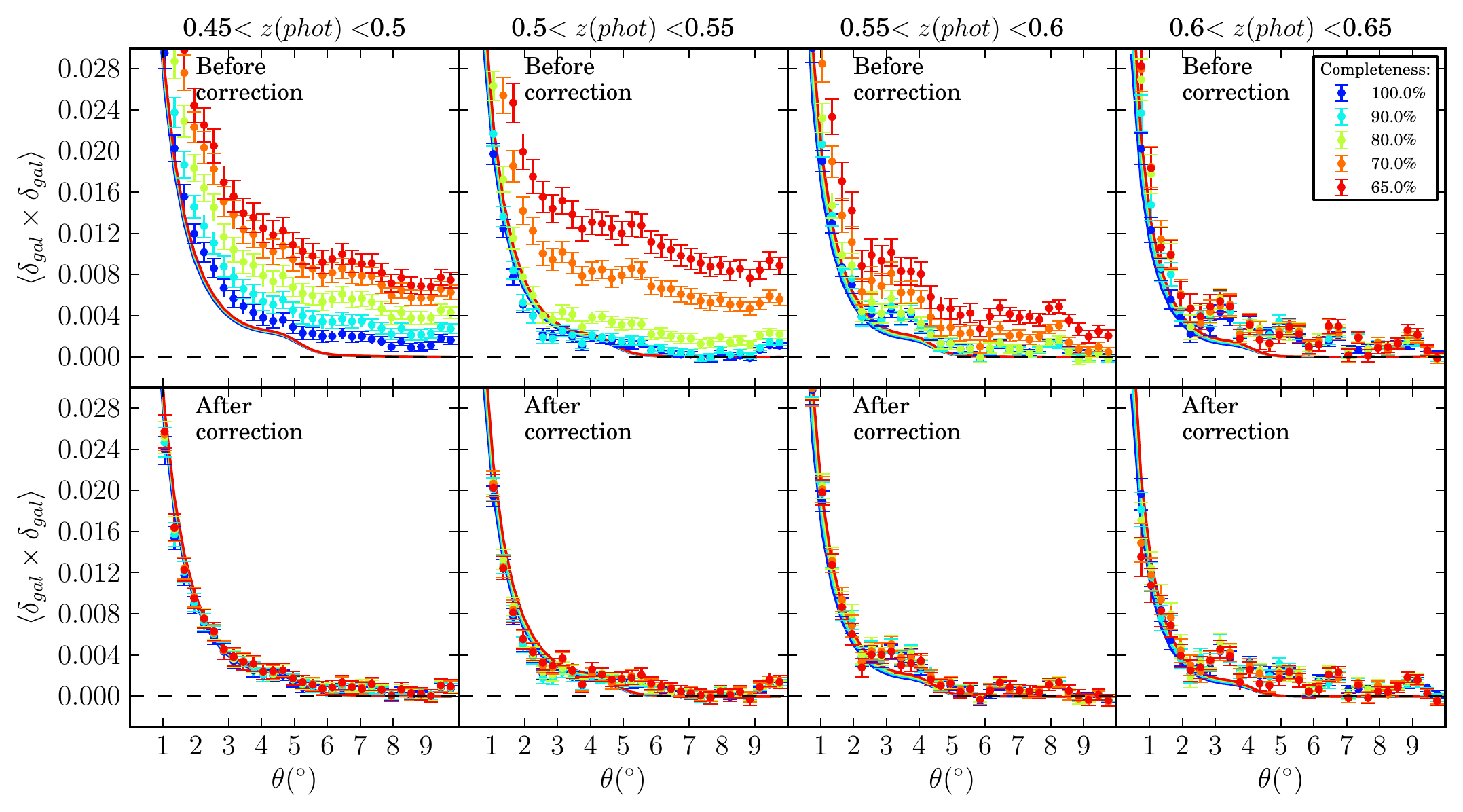}
\caption{The angular galaxy auto correlations in the four photo-z bins of Fig.~\ref{Nz_bins} with different photo-z quality cuts labeled with colors. The upper plots show the results before applying the \textit{odds} correction in (\ref{od_correction}), while the lower plots show the results after applying it. Points with error bars correspond to measurements and curves to predictions obtained using the $N_i(z)$ selection functions in Fig.~\ref{Nz_bins} in 
Eq.~(\ref{corr_prediction}).}
\label{auto_odcorr}
\end{figure*}

\begin{figure*}
\centering
\includegraphics[width=140mm]{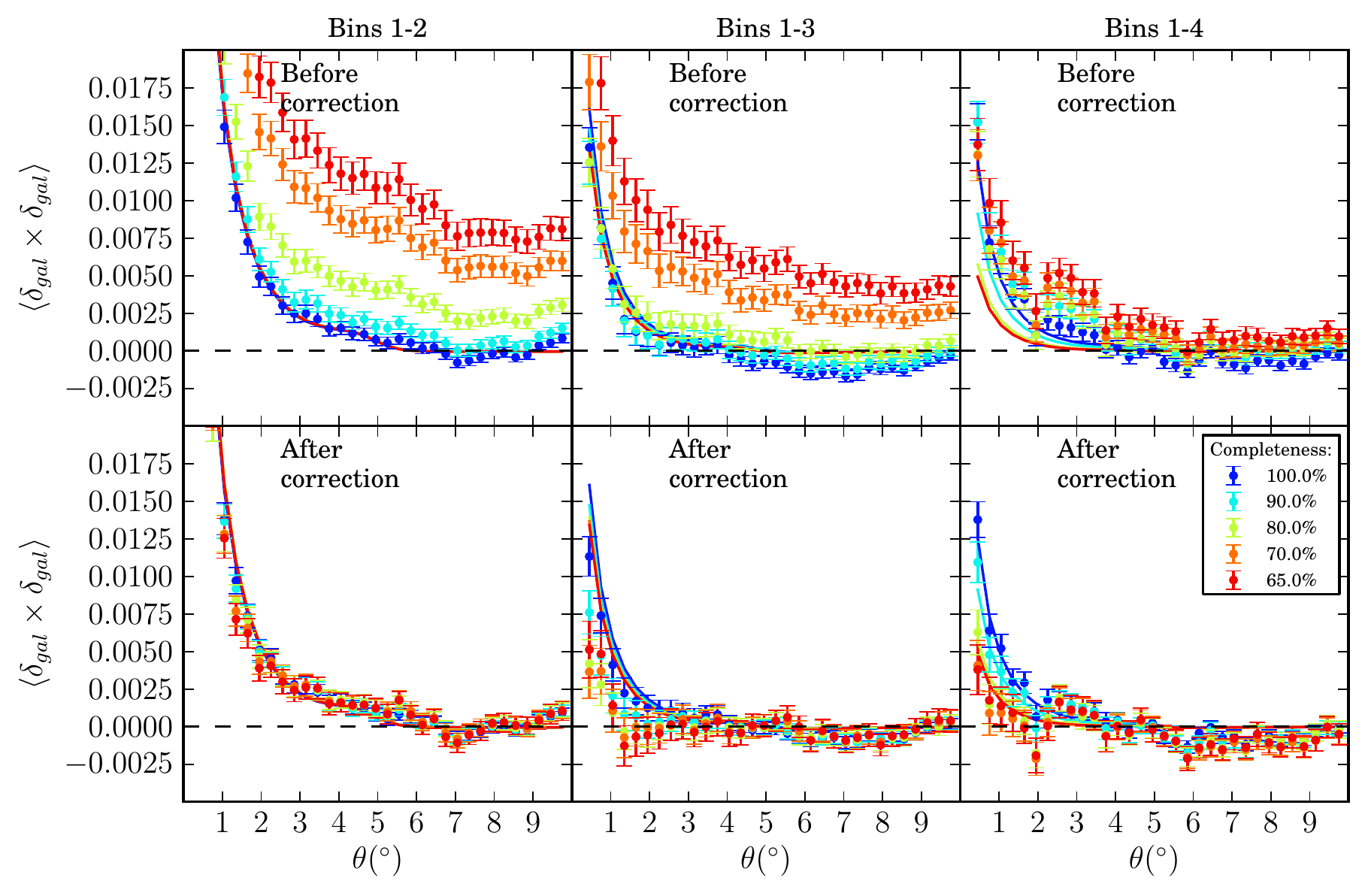} \\
\includegraphics[width=140mm]{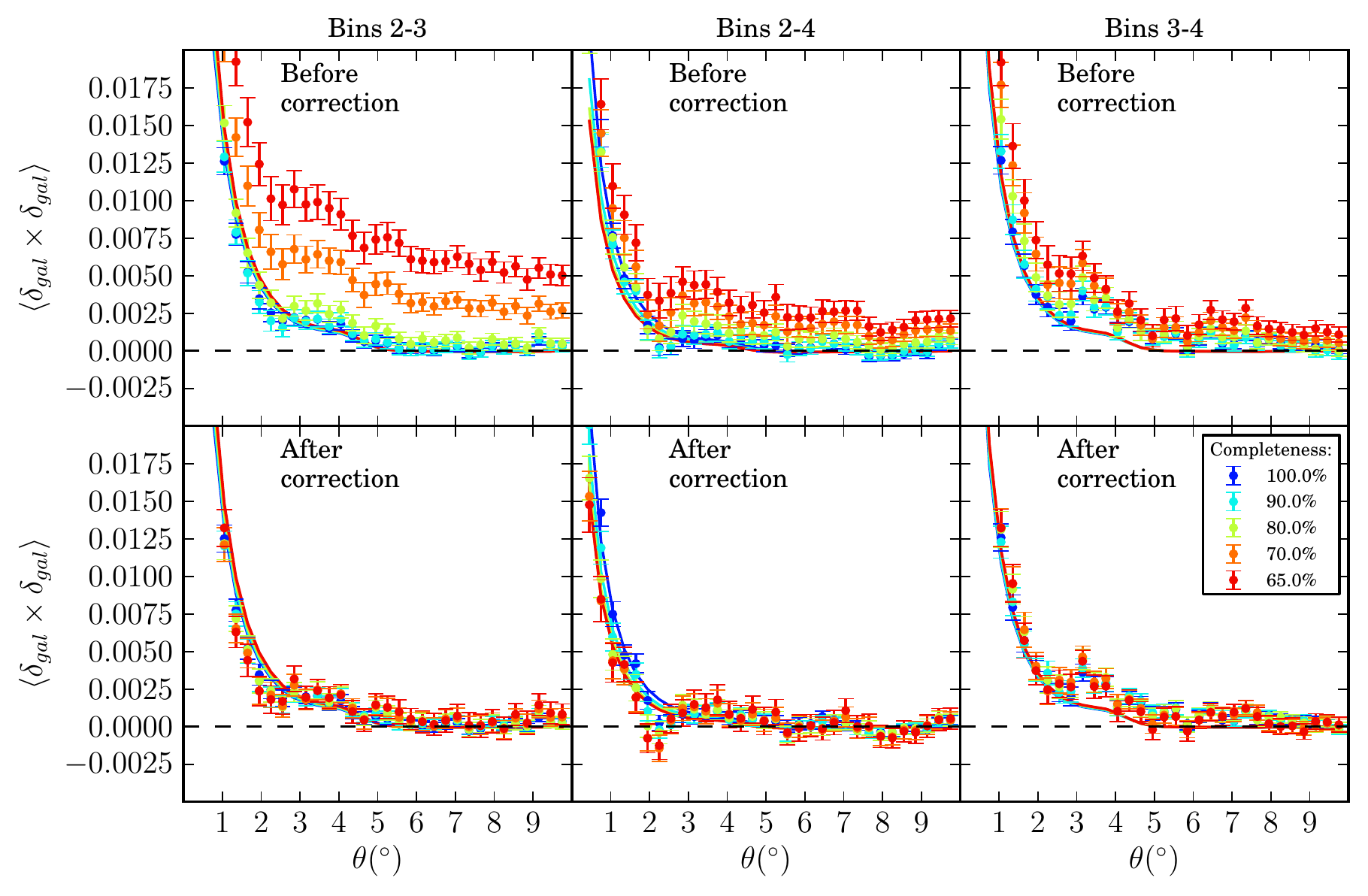}
\caption{All the possible combinations of the angular galaxy cross correlations between the different photo-z bins of Fig.~\ref{Nz_bins} with different photo-z quality cuts labeled with different colors. The upper plots show the results before applying the \textit{odds} correction in~(\ref{12od_correction}), while the lower plots show the results after applying it. Points with error bars correspond to measurements and curves to predictions obtained using the $N(z)$ selections functions in Fig.~\ref{Nz_bins} in Eq.~(\ref{corr_prediction}).}
\label{cross_odcorr}
\end{figure*}

The results for the auto-correlations are on the top row of Fig.~\ref{auto_odcorr}, and those for the cross-correlations are on the top row of Fig.~\ref{cross_odcorr}. Solid lines represent the predicted correlations obtained with (\ref{corr_prediction}) and points with error bars the measurements using (\ref{measured_correlations}) and (\ref{covariance}), respectively. Different colors label different photo-z quality cuts.

Focusing on the auto-correlations, we see that the results before any cut (blue), are slightly above the predicted curves (roughly 1 or $2\sigma$) in the first three photo-z bins, depending on the angle, and up to $\sim3\sigma$ in the last bin. The extra clustering in bin 0.6$<$z(phot)$<$0.65 is an issue already known from other studies such as \citet{Thomas:2010tn, Blake:2007xp}, and, in \citet{Crocce:2011mj}, it is justified by the fact that the number of galaxies in this bin is low enough for systematics to introduce significant distortions on the correlations. The photo-z quality cuts also introduce extra clustering, which is larger the harder the cut, as was seen in the galaxy-odds correlations in Fig.~\ref{gal_map}. Moreover, we see that it is also related to how much the odds are auto-correlated, since the increase of the correlations with the cut is lower at higher photo-z bins, which have smaller \textit{odds} auto-correlations (Fig.~\ref{od_map}). However, the correlations in the first bin do not increase as much as in the second bin, even having the most auto-correlated odds map. This might be related to the fact that galaxies in this bin are already correlated with the {\it odds} value before any cut, as we can see in the bottom-left plot of Fig.~\ref{gal_map}, and, in those cases, the extra clustering may not be as additive as for a completely uncorrelated clustering.

The cross-correlations show a similar behaviour, with the photo-z quality cuts introducing extra clustering. This effect is less significant in the cross-correlation of bins 1-4, 2-4 and 3-4 than in those of bins 1-3, 2-3 and 1-2. The reason is that the fourth bin is the one whose {\it odds} map has the lowest auto correlation.

\section{Correcting the effect of photo-z quality cuts on galaxy clustering}
\label{sec:correction}

In the previous section we have seen that the photo-z quality cuts introduce extra clustering in the angular galaxy correlations, since they remove galaxies non-homogeneously from the sky. In this section, we want to find a way to correct for it.

We will use the framework presented in~\citet{Ho:2012vy} and~\citet{Ross:2011cz}, which the authors use for the treatment of systematic effects that influence the SDSS-III galaxy clustering, such as the stellar contamination, the sky brightness or the image quality of the instrument. 
Following~\citet{Ho:2012vy} and~\citet{Ross:2011cz}, the density fluctuation $\delta_i$ of the systematic effect $i$ modifies the true galaxy density fluctuation $\delta^t_g$ through a linear contribution modulated by $\epsilon_i$, so that the observed galaxy density fluctuation becomes:
\begin{equation}
\delta_g = \delta^t_g+\sum_{i}\epsilon_i \delta_i \label{delta_g_obs} \, ,
\end{equation}
where the contribution of the systematic effect must be small compared with $\delta^t_g$.

Therefore, assuming that there is no intrinsic cross-correlation between the true galaxy fluctuations and the systematic effect, $\langle \delta^t_g \delta_i \rangle = 0$, the cross-correlation between the observed galaxy density fluctuation and the systematic effect is:
\begin{equation}
\langle \delta_g \delta_i \rangle = \langle (\delta^t_g+\sum_{j}\epsilon_j \delta_j) \delta_i \rangle = \epsilon_i \langle \delta_i \delta_i \rangle + \sum_{j\neq i}\epsilon_j\langle \delta_j \delta_i \rangle \label{omega_god} \, .
\end{equation}
If we consider that the only systematic effect acting is the \textit{odds} distribution, the previous equations reduce to:
\begin{eqnarray}
\delta_g = \delta^t_g+\epsilon_{od} \delta_{od} \label{delta_g} \\
\langle \delta^t_g \delta_{od} \rangle = 0 \label{god_zero_2} \\
\langle \delta_g \delta_{od} \rangle = \epsilon_{od}\langle \delta_{od} \delta_{od} \rangle \label{omega_god_2} \, .
\end{eqnarray}
Then, the angular galaxy cross-correlation between two different galaxy maps, 1 and 2, at, for instance, different redshifts, is:
\begin{align}
\langle \delta_{g1} \delta_{g2} \rangle &= \langle (\delta^t_{g1}+\epsilon_{od1} \delta_{od1}) (\delta^t_{g2}+\epsilon_{od2} \delta_{od2}) \rangle \nonumber \\
&= \langle \delta^t_{g1} \delta^t_{g2} \rangle + \epsilon_{od1} \epsilon_{od2} \langle \delta_{od1} \delta_{od2} \rangle \nonumber \\ 
&= \langle \delta^t_{g1} \delta^t_{g2} \rangle + {\langle \delta_{g1} \delta_{od1} \rangle \over \langle \delta_{od1} \delta_{od1} \rangle }{ \langle \delta_{g2} \delta_{od2} \rangle \over \langle \delta_{od2} \delta_{od2} \rangle} \langle \delta_{od1} \delta_{od2} \rangle \, ,
\end{align}
where in the first equality we have used~(\ref{delta_g}), in the second~(\ref{god_zero_2}) and in the third~(\ref{omega_god_2}). What we measure is the left-hand side of the equation, and, therefore, we need to subtract the second term on the right-hand side to obtain the true values of the correlations. 

Therefore, the \textit{odds} correction for the angular cross-correlations of two different galaxy maps is:
\begin{equation}
\omega^t_{g1,g2}(\theta) = \omega_{g1,g2}(\theta) - {\omega_{g1,od1}(\theta) \over \omega_{od1,od1}(\theta)}{\omega_{g2,od2}(\theta)\over\omega_{od2,od2}(\theta)}\omega_{od1,od2}(\theta) \, ,
\label{12od_correction}
\end{equation}
where $\omega^t_{g1,g2}(\theta)\equiv \langle \delta^t_{g1} \delta^t_{g2} \rangle_\theta$ is the true galaxy cross-correlation, $\omega_{g1,g2}(\theta) \equiv \langle \delta_{g1} \delta_{g2} \rangle_\theta$ is the observed one, $\omega_{g1,od1}(\theta) \equiv \langle \delta_{g1} \delta_{od1} \rangle_\theta$ is the cross-correlation of the galaxies with the \textit{odds} in map 1 (the same for map 2), $\omega_{od1,od1}(\theta) \equiv \langle \delta_{od1} \delta_{od1} \rangle_\theta$ is the auto-correlation of the \textit{odds} in  map 1 (the same for map 2), and $\omega_{od1,od2}(\theta) \equiv \langle \delta_{od1} \delta_{od2} \rangle_\theta$ is the cross-correlation between the \textit{odds} maps 1 and 2.

If we were only interested in this correction for auto-correlations
Eq.~(\ref{12od_correction}) would reduce to:
\begin{equation}
\omega^t_{g,g}(\theta) = \omega_{g,g}(\theta) - {\omega_{g,od}^2(\theta)\over\omega_{od,od}(\theta)} \, ,
\label{od_correction}
\end{equation}
where $\omega^t_{g,g}(\theta) \equiv \langle \delta^t_{g} \delta^t_{g} \rangle_\theta$ is the true galaxy auto-correlation, $\omega_{g,g}(\theta) \equiv \langle \delta_g \delta_g \rangle_\theta$ is the observed one, $\omega_{g,od}(\theta) \equiv \langle \delta_g \delta_{od} \rangle_\theta$ is the galaxy-odds cross-correlation, and $\omega_{od,od}(\theta) \equiv \langle \delta_{od} \delta_{od} \rangle_\theta$ is the odds auto-correlation.

The structure of Eqs.~(\ref{12od_correction}) and (\ref{od_correction}) is quite intuitive: we have the
correlations of the galaxies, shown on the top row of Figs.~\ref{auto_odcorr} and~\ref{cross_odcorr}, minus the cross-correlations of them with the photo-z quality, shown on the bottom row of~Fig.~\ref{gal_map}, properly normalized by the auto-correlations of the \textit{odds}.
Both $\omega_{g,g}(\theta)$ and $\omega_{g,od}(\theta)$ grow when the quality cuts are applied. Therefore, the increase in the auto-correlation will be compensated by the increase in the cross-correlation.

The origin of the galaxy-odds cross-correlation is probably manifold. On the one hand,  
the \textit{odds} map can be seen as a proxy for other systematic effects, such as sky brightness, seeing, airmass, etc, and correcting for it, even without any photo-z quality cut, could therefore partially correct for these other systematic effects.
For this reason, we will also apply these corrections when no cut is applied.
On the other hand, in a galaxy catalog containing both early- and late-type galaxies (unlike Mega-Z, which contains almost exclusively early-type galaxies), the \textit{odds} corrections could also reflect the fact that early-type galaxies cluster more than late-type galaxies, and, at the same time, their photo-z quality tends to be better, thereby creating a cross-correlation between galaxy clustering and \textit{odds}. 

\begin{figure}
\centering
\includegraphics[width=80mm]{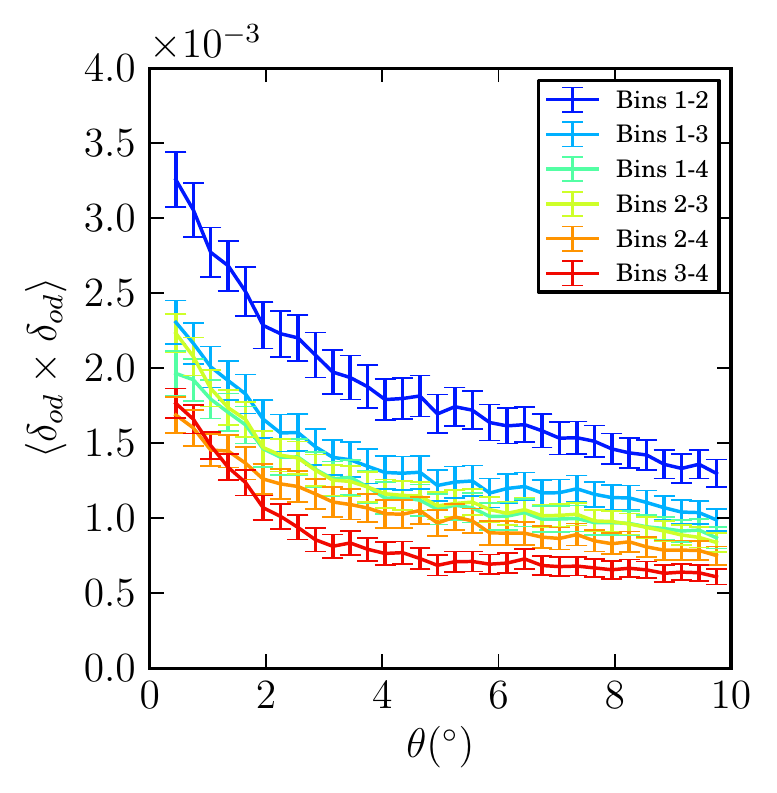}
\caption{All the possible combinations of the angular \textit{odds} cross-correlations between the different maps on the  top row of Fig~\ref{od_map}. They are needed for the \textit{odds} correction in~(\ref{12od_correction}). We see that all odds maps are cross-correlated with each other, at least up to angles $<10^\circ$. However, the higher the $z$, the smaller the correlation.}
\label{cross_od}
\end{figure}

In~Eq.~(\ref{12od_correction}) the cross-correlations between different \textit{odds} maps, $\langle \delta_{od1} \delta_{od2} \rangle$, are needed. In our case, those are the maps on the top row of Fig.~{\ref{od_map}}. We compute all the cross-scorrelations and display them on Fig.~\ref{cross_od}. We see that all the \textit{odds} maps are cross-correlated with each other, at least up to angles $<10^\circ$. However, the higher the $z$, the lower the correlation. For example, the bins 1-2 are the most cross-correlated, while the bins 3-4 are the least. All the other combinations have similar values.  

Finally, we apply the corrections to the measured angular galaxy clustering, before and after applying the photo-z quality cuts. 
The covariance $\text{Cov}_\omega(\theta_n , \theta_m )$ of the correlation functions after applying the \textit{odds} correction is obtained using Eq.~(\ref{covariance}) after applying the correction to each one of the jackknife correlation function $\omega_k(\theta_n)$.
The results are shown on the bottom rows of Fig.~\ref{auto_odcorr} for auto-correlations and Fig.~\ref{cross_odcorr} for cross-correlations. 
First, we see that the corrections work very well. After applying them, all measurements agree, regardless of the photo-z quality cut used in the analysis. 
Second, we see that the corrections not only work to remove the extra clustering introduced by the quality cuts, they also correct for any intrinsic extra clustering that might be present before any cut. For example, on the left-most plot of Fig.~\ref{auto_odcorr}, the auto-correlations were 1 or 2$\sigma$ above the prediction before any cut (blue). 
After applying the correction, most of the data points agree with their corresponding prediction.

The predictions in Fig.~\ref{auto_odcorr}, and most of those in Fig.~\ref{cross_odcorr} do not change much after the quality cuts, since they depend through (\ref{corr_prediction}) on the $N_i(z)$ functions in Fig.~\ref{Nz_bins},
which themselves change very little. In fact, the size of the measured error bars are not small enough to distinguish between different predictions. So much so that, although we have been able to correct for the extra clustering introduced by the quality cuts, the cuts themselves do not help in any relevant way in the clustering analysis, and, therefore, in this case there may be no obvious advantage in applying them. 
Furthermore, the relative errors in the corrected correlation functions can be substantially larger than those in the uncorrected ones: 
from a few percent larger to almost twice as large, depending on the angular scale, the photo-z bin and the value of the {\em odds} cut.

Things are different for cross-correlations. The strength of the signal of the cross-correlation between two different photo-z bins is mainly given by the amount of overlap in their $N_i(z)$, or, in other words, the fraction of galaxies that are at very similar true redshifts but, due to their photo-z uncertainty, end up in separate photo-z bins. In Fig.~\ref{Nz_bins}, we saw that the low tail of bin $0.6<z<0.65$, reduces considerably when the photo-z quality cuts are applied. Consequently, the overlap between this bin and the rest will also reduce, particularly the overlap with the farthest bin, $0.45<z<0.5$. This should result in differences in the predicted cross-correlations large enough to be distinguished by our measurements. If we look at the cross-correlation of the bins 1-4 on the top-right plot of Fig.~\ref{cross_odcorr}, we see that, at angles $<3^\circ$ where cross-correlations are not zero, the predicted curves differ more than the size of the error bars in the measurements. Even then, the corrections again put the measurements on top of their corresponding predicted curves. Note that, as for the auto-correlations, this is also true even when no {\em odds} cut is applied. This may have consequences for methods of photo-z calibration based on the study of the cross-correlations between photometric galaxy samples~\citep{Benjamin:2010rp}.

\section{Effects on the extraction of the BAO scale}
\label{sec:BAO}
Having studied in the previous sections the effects of photo-z quality cuts on the measured galaxy auto- and cross-correlations and the way to correct for them, next we want to see how the extraction of the Baryon Acoustic Oscillations (BAO) scale from the measured galaxy auto-correlations in Fig.~\ref{auto_odcorr} is affected by the 
photo-z quality cuts and the subsequent correction. 

The BAO scale has been proven to be a successful standard ruler to constrain cosmological parameters \citep{Eisenstein:2005su, Huetsi:2005tp, Percival:2007yw, Padmanabhan:2006cia, Okumura:2007br, Gaztanaga:2008de}. It was originated when primordial overdensities on matter caused acoustic (pressure) waves in the photon-baryon fluid that traveled freely across space until photons and baryons decoupled at the drag epoch (when the baryons were released from the ``Compton drag'' of the photons \citep{1996ApJ...471..542H}) $z_d=1059.25\pm0.58$ \citep{Ade:2013zuv}, and those waves stopped traveling. At that time, they had traveled $r_s(z_d)=147.49\pm0.59$~Mpc \citep{Ade:2013zuv} away from the primordial overdensities, where $r_s(z_d)$ is the sound horizon scale at the drag epoch. Later, when structure started forming, these waves seeded the formation of galaxies resulting in a small excess in the two-point angular galaxy clustering at the correspondent angular scale: 
\begin{equation}
\theta_{BAO}(z) = r_s (z_d)/ r(z) \, ,
\label{bao_scale}
\end{equation}
where
\begin{equation}
r(z) = \int^z_0 {dz \over H_0\sqrt{\Omega_M(1+z)^3+\Omega_\Lambda}}
\label{comov_dist}
\end{equation}
is the comoving distance at redshift $z$ that only depends on the Hubble constant $H_0=67.3\pm1.2$~(km/s)/Mpc and the fraction of matter $\Omega_M=0.315^{+0.016}_{-0.018}$ \citep{Ade:2013zuv} in a flat $\Lambda$CDM cosmological model. In this case $\Omega_\Lambda = 1 - \Omega_M =0.685^{+0.018}_{-0.016}$. 

A first detection and measurement of the BAO scale $\theta_{BAO}$ in angular clustering was presented in~\citet{Carnero:2011pu}. They made use of a method described in~\citet{Sanchez:2010zg} that consists of fitting the empirical function
\begin{equation}
\omega(\theta) = A + B\theta^\gamma + Ce^{-(\theta-\theta_{FIT})^2 /2\sigma^2}
\label{bao_fit}
\end{equation}
to the observed angular correlation function in a range of angular separations that encloses the BAO peak, where $\lbrace A,B,\gamma,C,\theta_{FIT},\sigma \rbrace$ are the parameters of the fit. $A$ takes into account any possible global offset, $B$ and $\gamma$ describe the typical decreasing profile of $\omega(\theta)$, with $\gamma$ negative, and the rest is a Gaussian that characterizes the BAO peak. A non-zero value of $C$ will tell us that the BAO peak has been detected, $\theta_{FIT}$ is the location of that peak and $\sigma$ its width. Since angular correlations are measured in redshift bins of finite width due to the intrinsic photo-z dispersion, projection effects can cause a mismatch between $\theta_{FIT}$ and $\theta_{BAO}$. \citet{Sanchez:2010zg} shows that this can be successfully corrected as follows:
\begin{equation}
\theta^{(obs)}_{BAO} = \alpha(z,\Delta z) \theta_{FIT} \, ,
\end{equation}
where $\alpha$ is a factor that only depends on the mean redshift $z$ and the width $\Delta z$ of the bin. Figure~3 in~\citet{Sanchez:2010zg} shows these dependences. In general terms, the wider the redshift bin, the larger the shift of $\theta_{FIT}$ towards smaller angles. For example, the shift is $\sim$5\% of $\theta_{BAO}$ when $\Delta z \sim 0.05$. For higher widths, only bins at $z\gtrsim0.5$ continue shifting, while the others saturate. 

In section~\ref{sec:photoz}, we chose a width of 0.05 for our photo-z bins in Fig.~\ref{Nz_bins}. This was in photo-z space. In real (spectroscopic) redshift space, this translates into a FWHM of $\sim$0.08, except in the last bin $0.6<z<0.65$ where, as a consequence of the poorer photo-z performance, the width becomes $\sim$0.12. We find that for those bins the shift in $\theta_{FIT}$ amounts to $\sim$7\%, except in the last bin, where it is $\sim$8\%.

\begin{figure*}
\centering
\includegraphics[type=pdf,ext=.pdf,read=.pdf, width=130mm]{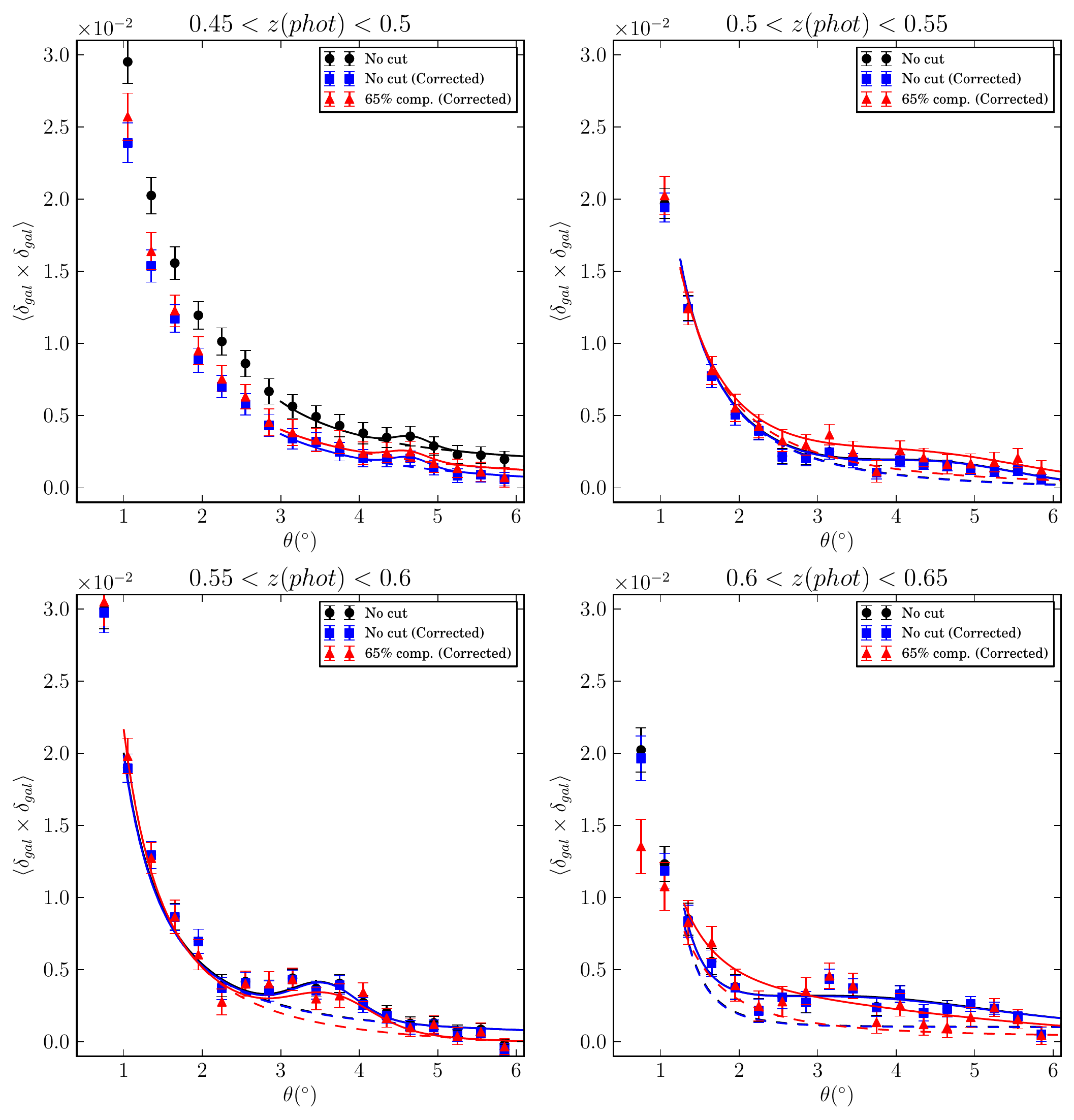}
\caption{Results of the fits of (\ref{bao_fit}) to the galaxy auto-correlations of Fig.~\ref{auto_odcorr} to extract the BAO scale. In black when no photo-z quality cut is applied, in blue the same but correcting for \textit{odds} and, in red when the 65\% completeness cut is applied and corrected. Error bars correspond to observations. Solid and dashed lines correspond both to the best fit, but in the dashed the BAO peak has been removed by setting $C=0$ in~(\ref{bao_fit}). 
}
\label{bao_4bins}
\end{figure*}

Finally, we fit (\ref{bao_fit}) to the galaxy auto-correlation measurements of Fig.~\ref{auto_odcorr}. The results are shown on Fig.~\ref{bao_4bins}. Fits are performed in three different cases: when no photo-z quality cut and odds correction are applied (black), when no photo-z quality cut is applied but the correction is (blue), and when the photo-z quality cut of 65\% completeness and the correction are applied (red). Error bars correspond to observations and are the same as in Fig.~\ref{auto_odcorr}. 
Solid and dashed lines correspond to the best fit, but in the dashed the BAO feature has been removed by setting $C=0$. The range of the fit is slightly changed in each photo-z bin to make sure that it completely encloses the BAO peak.

The results of the fits are summarized in Table~\ref{tab:BAO}. Looking at the parameter $C$, we see that we find evidence for the BAO peak (a non-zero value of $C$) in the first three bins, although with different significances in different bins. Typically, applying the photo-z quality cut and its correction (last column in Table~\ref{tab:BAO}) results in a decrease of the significance of the BAO peak, since 35\% of the galaxies are removed from the sample. The first and second column both correspond to the case in which no photo-z quality cut is applied. However, for the results in the second column the {\it odds} correction has been nevertheless applied using the formulas 
in~(\ref{od_correction}). This will correct for any intrinsic correlation between {\it odds} value and galaxy position that may create a spurious correlation in the galaxy map. Since higher {\it odds} may be due to larger signal-to-noise, the {\it odds} value may be seen as a proxy for airmass, seeing, extinction, etc. 
Therefore, it may make sense to correct for the {\it odds} effect even when applying no explicit {\it odds} cut in order to try to mitigate these effects. A comparison of the results in the second and third columns in Table~\ref{tab:BAO} shows that correcting for \textit{odds} when no cut is applied gives very similar results to simply not applying the correction. 
This is consistent with the findings in~\citet{Ho:2012vy} and~\citet{Ross:2011cz}, where they fail to see significant systematic effects in the galaxy auto-correlations due to differences in seeing, survey depth, extinction, etc. 

\begin{table*}
\caption{
Results of the BAO fits for the four photo-z bins, under three different conditions: no {\it odds} cut and no correction, no {\it odds} cut but correction for the {\it odds} effect, and finally after applying an {\it odds} cut that retains 65\% of the galaxies and the corresponding correction. The fit parameters are defined in Eq.~(\ref{bao_fit}). The rows labeled 
$\theta^{theo}_{BAO}$ contain the position of the BAO peak at the mean redshift in each bin derived from the latest Planck results~\citep{Ade:2013zuv}. The corresponding errors are derived by propagating the errors in the cosmological parameters in (\ref{bao_scale}) and (\ref{comov_dist}). Since the quality cut slightly changes the true redshift distribution inside each photo-z bin, the expected BAO scale in the bin may also change slightly.
} 
\vspace*{12pt}
\centering
\begin{tabular}{ c | c | c | c |}
\cline{2-4} 
 & \parbox{3cm}{\centering No odds cut} &  \parbox{3cm}{\centering No odds cut\\+ Correction} & \parbox{3cm}{\centering Odds cut (65\% eff.)\\+ Correction} \\ \cline{2-4}
\hhline{-===}
\multicolumn{4}{|c|}{$0.45<z<0.5$} \\ 
\hline
\multicolumn{1}{|c|}{C}& (0.7$\pm$0.3)$\cdot 10^{-3}$& (0.7$\pm$0.3)$\cdot 10^{-3}$& (0.6$\pm$0.4)$\cdot 10^{-3}$\\
\multicolumn{1}{|c|}{$\theta_{FIT}$}& 4.68$\pm$0.09& 4.65$\pm$0.09& 4.62$\pm$0.12\\
\multicolumn{1}{|c|}{$\theta^{obs}_{BAO}$}& 5.03$\pm$0.09& 5.00$\pm$0.08& 4.97$\pm$0.12\\
\multicolumn{1}{|c|}{$\theta^{theo}_{BAO}$}& 4.51$\pm$0.08& 4.51$\pm$0.08& 4.50$\pm$0.08\\
\hline
\hline
\multicolumn{4}{|c|}{$0.5<z<0.55$} \\ 
\hline
\multicolumn{1}{|c|}{C}& (1.2$\pm$0.4)$\cdot 10^{-3}$& (1.2$\pm$0.4)$\cdot 10^{-3}$& (1.5$\pm$0.6)$\cdot 10^{-3}$\\
\multicolumn{1}{|c|}{$\theta_{FIT}$}& 4.58$\pm$0.33& 4.59$\pm$0.32& 4.40$\pm$0.60\\
\multicolumn{1}{|c|}{$\theta^{obs}_{BAO}$}& 4.92$\pm$0.27& 4.94$\pm$0.27& 4.73$\pm$0.41\\
\multicolumn{1}{|c|}{$\theta^{theo}_{BAO}$}& 4.16$\pm$0.08& 4.16$\pm$0.08& 4.16$\pm$0.08\\
\hline
\hline
\multicolumn{4}{|c|}{$0.55<z<0.6$} \\ 
\hline
\multicolumn{1}{|c|}{C}& (2.2$\pm$0.4)$\cdot 10^{-3}$& (2.2$\pm$0.4)$\cdot 10^{-3}$& (2.2$\pm$0.6)$\cdot 10^{-3}$\\
\multicolumn{1}{|c|}{$\theta_{FIT}$}& 3.56$\pm$0.06& 3.57$\pm$0.06& 3.63$\pm$0.13\\
\multicolumn{1}{|c|}{$\theta^{obs}_{BAO}$}& 3.83$\pm$0.06& 3.84$\pm$0.06& 3.90$\pm$0.10\\
\multicolumn{1}{|c|}{$\theta^{theo}_{BAO}$}& 3.88$\pm$0.07& 3.88$\pm$0.07& 3.88$\pm$0.07\\
\hline
\hline
\multicolumn{4}{|c|}{$0.6<z<0.65$} \\ 
\hline
\multicolumn{1}{|c|}{C}& (2.1$\pm$0.5)$\cdot 10^{-3}$& (2.0$\pm$0.5)$\cdot 10^{-3}$& (2.0$\pm$4.4)$\cdot 10^{-3}$\\
\multicolumn{1}{|c|}{$\theta_{FIT}$}& 3.34$\pm$0.72& 3.30$\pm$0.77& 1.90$\pm$7.35\\
\multicolumn{1}{|c|}{$\theta^{obs}_{BAO}$}& 3.63$\pm$0.77& 3.59$\pm$0.81& 2.07$\pm$4.29\\
\multicolumn{1}{|c|}{$\theta^{theo}_{BAO}$}& 3.75$\pm$0.07& 3.75$\pm$0.07& 3.64$\pm$0.07\\
\hline
\end{tabular}
\label{tab:BAO}
\end{table*}

After applying the quality cut and its correction (last column), we recover values of $\theta^{obs}_{BAO}$ that are fully consistent with those found without applying an {\it odds} cut (first and second columns), demonstrating that our correction is consistent. In the last photo-z bin ($0.60 < z_{phot} < 0.65$), however,
the photo-z quality cut wipes out the BAO peak. We attribute this to the fact, already mentioned in section~\ref{sec:clustering}, that in this bin there are fewer galaxies than in the others, and even fewer after removing 35\% of them with the quality cut, and moreover, the photo-z performance is worse in this bin. 

It is worth mentioning that for some photo-z bins the theoretically-expected position of the BAO peak at the mean redshift in that photo-z bin, $\theta^{theo}_{BAO}$, changes slightly when the photo-z quality cut is applied. This is because $\theta^{theo}_{BAO}$ depends on $z$ (see Eq.~(\ref{bao_scale})) and, since the selection functions $N(z)$ change when quality cuts are applied, the mean redshift may also shift a little, shifting also $\theta_{BAO}$.  

We can see that in some bins there are some discrepancies between the extracted values of $\theta^{obs}_{BAO}$ and the expected $\theta^{theo}_{BAO}$. This is most likely due to systematic effects that lie beyond the scope of this paper. The errors quoted in Table~\ref{tab:BAO} only contain the statistical uncertainties from the fit.
Actually, the results of the fits are rather fragile, showing a significant dependence on details of the fits, such as the exact $\theta$ range chosen. Fits performed using the whole data covariance matrices (estimated with jack-knife) result in values for the $\chi^2$ per degree of freedom of order 2, signaling a poor fit quality. The same fits performed using only the diagonal elements of the covariance matrices result in very similar central values for $\theta_{FIT}$, but with larger errors and values of the $\chi^2$ per degree of freedom around 1. The main point of this section, however, can be qualitatively understood from looking only at the data points in Fig.~\ref{bao_4bins},  regardless of the fits: the position of the BAO peak does not change when going from data without {\em odds} cut and without correction to data without {\em odds} cut but with correction, and then to data with {\em odds} cut and with correction.

Finally, we have also tried to extract the BAO scale from the uncorrected auto-correlation functions after applying the most stringent {\em odds} cut. While in some bins, the fit has trouble converging, in those in which it does, the resulting BAO scale is only biased by a few percent, proving again the known fact that the BAO scale is very robust against systematic errors, even those, like this one, that grossly bias the overall shape and normalization of the correlation function. For example, in the bin with $0.55 < z < 0.60$, the BAO peak is found at $\theta^{obs}_{BAO} = (3.91 \pm 0.09)$~deg when applying the {\em odds} cut and no correction, to be compared with $\theta^{obs}_{BAO} = (3.90 \pm 0.10)$~deg, obtained applying the correction.

The main result of this section, however, is not another measurement of the BAO scale with the SDSS LRG sample, but rather the proof that after applying a tight photo-z quality cut that eliminates 35\% of the galaxies and severely distorts the shape of the galaxy-galaxy auto-correlation, the correction technique outlined in the previous section delivers a corrected auto-correlation function from which the BAO feature can be extracted without introducting any additional bias.

\section{Discussion and Conclusions}
\label{sec:discussion}
In the previous sections we have seen how photo-z
quality cuts, if left uncorrected, can severely bias the measured
galaxy angular auto- and cross-correlations. The effect, as seen in
Figs.~\ref{auto_odcorr} and~\ref{cross_odcorr}, consists mostly of a
large increase in the correlation across the whole range in angular
separation, although slightly more prominent at larger
separations. This is not unlike the effect reported in other clustering
studies based on very similar samples, such as those in~\citet{Thomas:2010tn}
and in~\citet{Crocce:2011mj}. In those papers an excess of clustering has
been observed in the photometric redshift bin $0.6 < z < 0.65$. In at
least one of the papers~\citep{Crocce:2011mj} a photo-z quality cut is
perfomed, eliminating about 16\% of the galaxies, but no attempt is
made to correct for the possible effect of this cut on the measured
correlations.  While a quick look at Fig.~\ref{auto_odcorr} reveals
that in our case the effect of the photo-z quality cut is more
prominent at lower redshifts, the issue may deserve more thorough study, 
which lies beyond the scope of this paper.

It is intriguing to see that, at least in the $0.45 < z < 0.50$
photo-z bin, there is a correlation between galaxy density and 
{\em odds} value even before any cut on the value of the {\em odds}
(bottom-left plot in Fig.~\ref{gal_map}). This
correlation then leads to extra galaxy auto- and cross-correlations
whenever that first photo-z bin is involved (top-left plot in
Fig.~\ref{auto_odcorr} and top-right plot in Fig.~\ref{cross_odcorr}). 
The correction method we propose
eliminates this extra correlation very effectively (see the corresponding plots in
Figs.~\ref{auto_odcorr} and \ref{cross_odcorr}), but the question
remains: what is it that we are actually eliminating? Or: where does
this galaxy-{\em odds} correlation come from? 

One possibility is that
it comes from systematic effects in the survey that are otherwise
uncorrected: differences in seeing conditions, airmass, extinction,
etc. between different areas of the survey will lead to correlated differences in
galaxy density and in the value of the {\em odds} parameter. In
general, these systematic effects would result in an additive extra
correlation. In this case, correcting for this
spurious correlation will mitigate the effects of those systematic issues.

However, in general, another possibility is that the galaxy-{\em odds} correlation is due to the
fact that different galaxy types have different clustering amplitudes
(different biases) and, at the same time, also have different mean
photo-z precisions, hence different mean {\em odds}. 
In this case, the {\em odds} correction could be removing genuine
galaxy-galaxy correlations. Alternatively, one could interpret that
the correction would be changing the average galaxy
type (and hence bias) of the sample. Hence, this would result in a
multiplicative extra correlation.

In our case, since the sample is largely composed of LRGs, the
galaxy-{\em odds}  correlation we observe in the lower photo-z bin is likely 
due to the uncorrected systematic effects mentioned above, and,
therefore, it makes sense to apply the {\em odds} correction even without an
explicit {\em odds} cut. We observe that, indeed, the extra
correlation we observe seems to be additive in nature (i.e.~roughly
constant as a function of angular scale).
Figures~\ref{auto_odcorr} and \ref{cross_odcorr} show that, even with
no {\em odds} cut, the agreement
with the predictions improves once the {\em odds} correction
has been applied.

\vspace*{1em}

In summary, using the Mega-Z DR7 galaxy sample and the {\tt BPZ}
photometric redshift code,
we have shown that applying moderate galaxy photo-z quality cuts
may lead to large biases in the measured galaxy auto- and cross-correlations.
However, a correction method
derived within the framework presented in~\citet{Ho:2012vy, Ross:2011cz}
manages to recover the original correlation functions and,
in particular, does not bias the extraction of the BAO peak. It
remains to be seen whether this correction might eliminate some or
all of the excess correlation observed by several groups in galaxy
samples essentially identical to the one we use.

\section*{Acknowledgments} 
We thank Linda \"Ostman for her help in the early stages of this work.
Funding for this project was partially provided by the Spanish
Ministerio de Ciencia e Innovaci\'on under projects
AYA2009-13936-C06-01 and -02, FPA2012-39684-C03-01,
and Consolider-Ingenio CSD2007-00060. 

\bibliography{mnras_paper}

\begin{thebibliography}{}

\bibitem[\protect\citeauthoryear{Ade et~al.,}{Ade  et~al.}{2013}]{Ade:2013zuv}
Ade P.,  et~al., 2013, arXiv:1303.5076

\bibitem[\protect\citeauthoryear{Arnouts, Cristiani, Moscardini, Matarrese,
  Lucchin et~al.,}{Arnouts et~al.}{1999}]{Arnouts:1999bb}
Arnouts S.,  Cristiani S.,  Moscardini L.,  Matarrese S.,  Lucchin F.,
  et~al., 1999, MNRAS, 310, 540

\bibitem[\protect\citeauthoryear{Benitez}{Benitez}{2000}]{Benitez:1998br}
Benitez N.,  2000, ApJ, 536, 571

\bibitem[\protect\citeauthoryear{Benitez, Gaztanaga, Miquel, Castander, Moles
  et~al.,}{Benitez et~al.}{2009}]{Benitez:2008fs}
Benitez N.,  Gaztanaga E.,  Miquel R.,  Castander F.,  Moles M.,    et~al.,
  2009, ApJ, 691, 241

\bibitem[\protect\citeauthoryear{Benjamin, Van~Waerbeke, Menard \&
  Kilbinger}{Benjamin et~al.}{2010}]{Benjamin:2010rp}
Benjamin J.,  Van~Waerbeke L.,  Menard B.,    Kilbinger M.,  2010, MNRAS, 408,
  1168

\bibitem[\protect\citeauthoryear{Blake, Collister \& Lahav}{Blake
  et~al.}{2008}]{Blake:2007xp}
Blake C.,  Collister A.,    Lahav O.,  2008, MNRAS, 385, 1257

\bibitem[\protect\citeauthoryear{Bolzonella, Miralles \& Pello'}{Bolzonella
  et~al.}{2000}]{Bolzonella:2000js}
Bolzonella M.,  Miralles J.-M.,    Pello' R.,  2000, A\&A, 363, 476

\bibitem[\protect\citeauthoryear{{Brammer}, {van Dokkum} \& {Coppi}}{{Brammer}
  et~al.}{2008}]{2008ApJ...686.1503B}
{Brammer} G.~B.,  {van Dokkum} P.~G.,    {Coppi} P.,  2008, ApJ, 686, 1503

\bibitem[\protect\citeauthoryear{Cabre, Fosalba, Gaztanaga \& Manera}{Cabre
  et~al.}{2007}]{Cabre:2007rv}
Cabre A.,  Fosalba P.,  Gaztanaga E.,    Manera M.,  2007, MNRAS, 381, 1347

\bibitem[\protect\citeauthoryear{Cabre \& Gaztanaga}{Cabre \&
  Gaztanaga}{2009}]{Cabre:2008sz}
Cabre A.,  Gaztanaga E.,  2009, MNRAS, 393, 1183

\bibitem[\protect\citeauthoryear{Cannon, Drinkwater, Edge, Eisenstein, Nichol
  et~al.,}{Cannon et~al.}{2006}]{Cannon:2006qh}
Cannon R.,  Drinkwater M.,  Edge A.,  Eisenstein D.,  Nichol R.,    et~al.,
  2006, MNRAS, 372, 425

\bibitem[\protect\citeauthoryear{Carnero, Sanchez, Crocce, Cabre \&
  Gaztanaga}{Carnero et~al.}{2012}]{Carnero:2011pu}
Carnero A.,  Sanchez E.,  Crocce M.,  Cabre A.,    Gaztanaga E.,  2012, MNRAS,
  419, 1689

\bibitem[\protect\citeauthoryear{{Carrasco Kind} \& {Brunner}}{{Carrasco Kind}
  \& {Brunner}}{2013}]{2013MNRAS.432.1483C}
{Carrasco Kind} M.,  {Brunner} R.~J.,  2013, MNRAS, 432, 1483

\bibitem[\protect\citeauthoryear{Coe, Benitez, Sanchez, Jee, Bouwens
  et~al.,}{Coe et~al.}{2006}]{Coe:2006hj}
Coe D.,  Benitez N.,  Sanchez S.~F.,  Jee M.,  Bouwens R.,    et~al., 2006, AJ,
  132, 926

\bibitem[\protect\citeauthoryear{Coleman, Wu \& Weedman}{Coleman
  et~al.}{1980}]{Coleman:1980}
Coleman D.,  Wu C.,    Weedman D.,  1980, ApJ, 43, 393

\bibitem[\protect\citeauthoryear{Colless et~al.,}{Colless
  et~al.}{2001}]{Colless:2001gk}
Colless M.,  et~al., 2001, MNRAS, 328, 1039

\bibitem[\protect\citeauthoryear{Collister, Lahav, Blake, Cannon, Croom
  et~al.,}{Collister et~al.}{2007}]{Collister:2006qg}
Collister A.,  Lahav O.,  Blake C.,  Cannon R.,  Croom S.,    et~al., 2007,
  MNRAS, 375, 68

\bibitem[\protect\citeauthoryear{Collister \& Lahav}{Collister \&
  Lahav}{2004}]{Collister:2003cz}
Collister A.~A.,  Lahav O.,  2004, Publ.Astron.Soc.Pac., 116, 345

\bibitem[\protect\citeauthoryear{Crocce, Cabre \& Gaztanaga}{Crocce
  et~al.}{2011}]{Crocce:2010qi}
Crocce M.,  Cabre A.,    Gaztanaga E.,  2011, MNRAS, 414, 329

\bibitem[\protect\citeauthoryear{Crocce, Gaztanaga, Cabre, Carnero \&
  Sanchez}{Crocce et~al.}{2011}]{Crocce:2011mj}
Crocce M.,  Gaztanaga E.,  Cabre A.,  Carnero A.,    Sanchez E.,  2011, MNRAS,
  417, 2577

\bibitem[\protect\citeauthoryear{Dawson et~al.,}{Dawson
  et~al.}{2013}]{2013AJ....145...10D}
Dawson K.~S.,  et~al., 2013, AJ, 145, 10

\bibitem[\protect\citeauthoryear{Drinkwater, Jurek, Blake, Woods, Pimbblet
  et~al.,}{Drinkwater et~al.}{2010}]{Drinkwater:2009sd}
Drinkwater M.~J.,  Jurek R.~J.,  Blake C.,  Woods D.,  Pimbblet K.~A.,
  et~al., 2010, MNRAS, 401, 1429

\bibitem[\protect\citeauthoryear{Efron}{Efron}{1979}]{efron79}
Efron B.,  1979, Ann. Statistics, 7, 1

\bibitem[\protect\citeauthoryear{Eisenstein et~al.,}{Eisenstein
  et~al.}{2005}]{Eisenstein:2005su}
Eisenstein D.~J.,  et~al., 2005, ApJ, 633, 560

\bibitem[\protect\citeauthoryear{Fukugita, Ichikawa, Gunn, Doi, Shimasaku
  et~al.,}{Fukugita et~al.}{1996}]{Fukugita:1996qt}
Fukugita M.,  Ichikawa T.,  Gunn J.,  Doi M.,  Shimasaku K.,    et~al., 1996,
  AJ, 111, 1748

\bibitem[\protect\citeauthoryear{Gaztanaga, Miquel \& Sanchez}{Gaztanaga
  et~al.}{2009}]{Gaztanaga:2008de}
Gaztanaga E.,  Miquel R.,    Sanchez E.,  2009, Phys.Rev.Lett., 103, 091302

\bibitem[\protect\citeauthoryear{Gerdes, Sypniewski, McKay, Hao, Weis
  et~al.,}{Gerdes et~al.}{2010}]{Gerdes:2009tw}
Gerdes D.~W.,  Sypniewski A.~J.,  McKay T.~A.,  Hao J.,  Weis M.~R.,    et~al.,
  2010, ApJ, 715, 823

\bibitem[\protect\citeauthoryear{Gorski, Hivon, Banday, Wandelt, Hansen
  et~al.,}{Gorski et~al.}{2005}]{Gorski:2004by}
Gorski K.,  Hivon E.,  Banday A.,  Wandelt B.,  Hansen F.,    et~al., 2005,
  ApJ, 622, 759

\bibitem[\protect\citeauthoryear{Ho, Cuesta, Seo, de Putter, Ross et~al.,}{Ho
  et~al.}{2012}]{Ho:2012vy}
Ho S.,  Cuesta A.,  Seo H.-J.,  de Putter R.,  Ross A.~J.,    et~al., 2012,
  ApJ, 761, 14

\bibitem[\protect\citeauthoryear{{Hu} \& {Sugiyama}}{{Hu} \&
  {Sugiyama}}{1996}]{1996ApJ...471..542H}
{Hu} W.,  {Sugiyama} N.,  1996, ApJ, 471, 542

\bibitem[\protect\citeauthoryear{Huetsi}{Huetsi}{2006}]{Huetsi:2005tp}
Huetsi G.,  2006, A\&A, 449, 891

\bibitem[\protect\citeauthoryear{Ilbert, Arnouts, McCracken, Bolzonella, Bertin
  et~al.,}{Ilbert et~al.}{2006}]{Ilbert:2006dp}
Ilbert O.,  Arnouts S.,  McCracken H.,  Bolzonella M.,  Bertin E.,    et~al.,
  2006, A\&A, 457, 841

\bibitem[\protect\citeauthoryear{Kaiser}{Kaiser}{1984}]{Kaiser:1984sw}
Kaiser N.,  1984, ApJ, 284, L9

\bibitem[\protect\citeauthoryear{Kaiser, Tonry \& Luppino}{Kaiser
  et~al.}{2000}]{2000PASP..112..768K}
Kaiser N.,  Tonry J.,    Luppino G.,  2000, pasp, 112, 768

\bibitem[\protect\citeauthoryear{Kinney, Calzetti, Bohlin, McQuade,
  Storchi-Bergmann et~al.,}{Kinney et~al.}{1996}]{Kinney:1996zz}
Kinney A.~L.,  Calzetti D.,  Bohlin R.~C.,  McQuade K.,  Storchi-Bergmann T.,
   et~al., 1996, ApJ, 467, 38

\bibitem[\protect\citeauthoryear{Le~Fevre, Vettolani, Garilli, Tresse, Bottini
  et~al.,}{Le~Fevre et~al.}{2005}]{LeFevre:2004hv}
Le~Fevre O.,  Vettolani G.,  Garilli B.,  Tresse L.,  Bottini D.,    et~al.,
  2005, A\&A, 439, 845

\bibitem[\protect\citeauthoryear{Mart\'{i}, Miquel, Castander
  et~al.,}{Mart\'{i} et~al.}{2014}]{Marti:2013}
Mart\'{i} P.,  Miquel R.,  Castander F.~J.,    et~al., 2014, In preparation

\bibitem[\protect\citeauthoryear{Okumura, Matsubara, Eisenstein, Kayo, Hikage
  et~al.,}{Okumura et~al.}{2008}]{Okumura:2007br}
Okumura T.,  Matsubara T.,  Eisenstein D.~J.,  Kayo I.,  Hikage C.,    et~al.,
  2008, ApJ, 676, 889

\bibitem[\protect\citeauthoryear{Padmanabhan et~al.,}{Padmanabhan
  et~al.}{2005}]{Padmanabhan:2004ic}
Padmanabhan N.,  et~al., 2005, MNRAS, 359, 237

\bibitem[\protect\citeauthoryear{Padmanabhan et~al.,}{Padmanabhan
  et~al.}{2007}]{Padmanabhan:2006cia}
Padmanabhan N.,  et~al., 2007, MNRAS, 378, 852

\bibitem[\protect\citeauthoryear{Percival, Cole, Eisenstein, Nichol, Peacock
  et~al.,}{Percival et~al.}{2007}]{Percival:2007yw}
Percival W.~J.,  Cole S.,  Eisenstein D.~J.,  Nichol R.~C.,  Peacock J.~A.,
  et~al., 2007, MNRAS, 381, 1053

\bibitem[\protect\citeauthoryear{Ross, Ho, Cuesta, Tojeiro, Percival
  et~al.,}{Ross et~al.}{2011}]{Ross:2011cz}
Ross A.~J.,  Ho S.,  Cuesta A.~J.,  Tojeiro R.,  Percival W.~J.,    et~al.,
  2011, MNRAS, 417, 1350

\bibitem[\protect\citeauthoryear{Sanchez, Carnero, Garcia-Bellido, Gaztanaga,
  de Simoni et~al.,}{Sanchez et~al.}{2011}]{Sanchez:2010zg}
Sanchez E.,  Carnero A.,  Garcia-Bellido J.,  Gaztanaga E.,  de Simoni F.,
  et~al., 2011, MNRAS, 411, 277

\bibitem[\protect\citeauthoryear{Smith et~al.,}{Smith
  et~al.}{2003}]{Smith:2002dz}
Smith R.,  et~al., 2003, MNRAS, 341, 1311

\bibitem[\protect\citeauthoryear{Thomas, Abdalla \& Lahav}{Thomas
  et~al.}{2011a}]{Thomas:2010tn}
Thomas S.~A.,  Abdalla F.~B.,    Lahav O.,  2011a, Phys.Rev.Lett., 106, 241301

\bibitem[\protect\citeauthoryear{Thomas, Abdalla \& Lahav}{Thomas
  et~al.}{2011b}]{Thomas:2011cu}
Thomas S.~A.,  Abdalla F.~B.,    Lahav O.,  2011b, MNRAS, 412, 1669

\bibitem[\protect\citeauthoryear{Tyson, Wittman, Hennawi \& Spergel}{Tyson
  et~al.}{2003}]{Tyson:2002nh}
Tyson J.,  Wittman D.,  Hennawi J.,    Spergel D.,  2003,
  Nucl.Phys.Proc.Suppl., 124, 21

\bibitem[\protect\citeauthoryear{York et~al.,}{York
  et~al.}{2000}]{York:2000gk}
York D.~G.,  et~al., 2000, AJ, 120, 1579

\end{thebibliography}

\label{lastpage}

\bibliographystyle{mn2e}

\end{document}